\documentstyle[editedvolume,numreferences,psfig,bezier,epsf]{crckapb} 

\input{prepictex.tex}
\input{pictex.tex}
\input{postpictex.tex}

\newcommand{\rf}[1]{(\ref{#1})}
\newcommand{\beq}{\begin{equation}}
\newcommand{\eeq}{\end{equation}}
\newcommand{\bea}{\begin{eqnarray}}
\newcommand{\eea}{\end{eqnarray}}

\newcommand{\e}{\mbox{e}}
\renewcommand{\d}{\mbox{d}}
\newcommand{\g}{\gamma}

\renewcommand{\l}{\lambda}
\renewcommand{\L}{\Lambda}

\renewcommand{\a}{\alpha}
\newcommand{\n}{\nu}  
\newcommand{\m}{\mu}

\newcommand{\ep}{\varepsilon}

\newcommand{\del}{\delta}

\newcommand{\k}{\kappa}

\newcommand{\oh}{\frac{1}{2}}

\newcommand{\ra}{\right\rangle}
\newcommand{\la}{\left\langle}
\newcommand{\prt}{\partial}
\newcommand{\mi}{\!-\!}
\newcommand{\equ}{\!=\!}
\newcommand{\pl}{\!+\!}

\newcommand{\cD}{{\cal D}}

\newcommand{\cT}{{\cal T}}

\newcommand{\cO}{{\cal O}}

\newcommand{\tL}{{\tilde{\L}}}
\newcommand{\tX}{{\tilde{X}}}
\newcommand{\tY}{{\tilde{Y}}}

\newcommand{\no}{\nonumber}
\newcommand{\nn}{\no\\}

\newcommand{\ointz}{\oint \frac{dz}{2\pi i \, z}\;}

\newcommand{\SL}{\sqrt{\L}}
\newcommand{\SLT}{\sqrt{\L}T}
\newcommand{\R}{{\rm I\!R}}

\begin{opening}

\title{Lorentzian and Euclidean Quantum Gravity -- 
Analytical and Numerical Results}

\subtitle{Lectures presented at the 1999 NATO-ASI on ``Quantum Geometry'' 
in Akureyri, Iceland. \\ 
Preprint: NBI-HE-00-02, AEI-2000-0
}

\author{J. AMBJ\O RN}
\institute{The Niels Bohr Institute, \\
Blegdamsvej 17, DK-2100 Copenhagen \O , Denmark\\
           email: ambjorn@nbi.dk}

\author{J. Jurkiewicz}
\institute{Institute of Physics, Jagellonian University, \\
ul. Reymonta 4, PL-30 059, Krak\'{o}w 16, Poland\\
email: jurkiewi@thrisc.if.uj.edu.pl}

\author{R. Loll}
\institute{Albert Einstein Institute\\
(Max-Planck-Institute for Gravitational Physics),\\
Am M\"{u}hlenberg 1, D-14476 Golm, Germany\\
email: loll@aei-potsdam.mpg.de}

\end{opening}

\runningtitle{Quantum Gravity}

\begin{document}
\begin{abstract}
We review some recent attempts to extract information about the
nature of quantum gravity, with and without matter,
by quantum field theoretical methods.
More specifically, we work within a covariant lattice approach
where the individual space-time geometries are constructed from
fundamental simplicial building blocks, and
the path integral over geometries is approximated by
summing over a class of piece-wise linear geometries.
This method of ``dynamical triangulations'' 
is very powerful in 2d, where the
regularized theory can be solved explicitly, and
gives us more insights into the quantum nature of 2d space-time
than continuum methods are presently able to provide. 
It also allows us to establish
an explicit relation between the Lorentzian- and Euclidean-signature
quantum theories. 
Analogous regularized gravitational models can be set up
in higher dimensions. Some analytic tools exist to study their
state sums, but, unlike in 2d, no complete analytic
solutions have yet been constructed.
However, a great advantage of our approach is the fact
that it is well-suited for numerical simulations. 
In the second part of this review we describe the
relevant Monte Carlo techniques, as well as some of 
the physical results that have been obtained from the simulations of
Euclidean gravity. We also explain why the Lorentzian version of dynamical
triangulations is a promising candidate for a non-perturbative
theory of quantum gravity.

\end{abstract}

\section{Introduction}
\label{sec:intro}

There is at present no satisfactory theory of four-dimensional 
quantum gravity. This is partly related to the
conceptual questions that arise when dealing with
fluctuating geometries without reference to any background 
metric. 
Addressing these seems to call for a genuinely non-perturbative 
formulation of quantum gravity. The need for a non-perturbative
approach persists in formulations 
where gravity appears embedded into a larger theory, such as
string or M-theory. Although there have been attempts to
identify non-perturbative structures within these unified
theories (see \cite{susskind,kawaietal} and many successive papers), 
they so far seem to have raised as many
questions as they have been able to answer. 
In these lectures we will review an alternative non-perturbative
approach to quantum gravity. It is more conservative in spirit,
in that it does not conjecture the existence of any radically new  
physical principles or symmetries.
This so-called ``dynamical triangulations'' approach\footnote{The method
of dynamical triangulations was introduced in the context of string 
theory and 2d quantum gravity in \cite{david,adf,kkm}, and 
subsequently extended to  
higher-dimensional Euclidean quantum gravity \cite{aj,am}. 
An extensive review covering the developments up to 1996 
can be found in the book \cite{book}. 
A more recent summary is contained in \cite{andre},
while the review \cite{l-review} deals with a variety of
lattice approaches to four-dimensional quantum gravity,
including dynamical triangulations. The use of dynamical-triangulations
methods in Lorentzian gravity was pioneered in \cite{al,aal1,aal2}.} can
be formulated entirely within the framework of ordinary quantum
field theory, while taking into account that the governing symmetry
principle is reparametrization- (diffeomorphism-) invariance.

Our ansatz can be made in any dimension, and has been solved
analytically in two space-time dimensions.
In higher dimensions one so far has to rely largely on numerical 
simulations, and up to now only the case of Euclidean signature
has been studied in detail. 
Although gravity in two dimensions has no genuine 
dynamical degrees of freedom, 
it does provide a testing ground for addressing some of the 
conceptual problems of quantum 
gravity explicitly, for example, how to define a notion of 
reparametrization-invariant length in a 
fluctuating geometry or how to define correlation functions
without reference to a background geometry. 
In 2d Euclidean gravity, a number of questions can be answered 
by continuum functional methods (within Liouville quantum field theory),
but some of the most interesting problems involving geodesic distances
can only be addressed in the non-perturbative setting of 
dynamical triangulations. The fact that in this latter approach
quantum gravity is obtained as the scaling limit of a lattice theory 
shows that, contrary to common belief, discrete methods can be
used also for reparametrization-invariant theories, as long as the 
discretization takes place directly on the space of geometries
(that is, on the space of metrics modulo diffeomorphisms). 

Another point that has been analyzed in the 
two-dimensional model is the analytic continuation 
between geometries of Lorentzian and Euclidean signature.
Canonical quantization attempts of (Lorentzian) gravity
usually take as their starting point globally hyperbolic
manifolds equipped with a non-degenerate Lorentzian metric
(which gives rise to a causal structure).
However, it is {\it a priori} unclear to what extent these essentially
classical structures should be preserved in the quantum theory.
For example, in a covariant path-integral approach it is
not obvious that a causality constraint should be imposed 
on each individual space-time configuration contributing to the 
amplitude. This idea was first advocated by Teitelboim 
\cite{teitelboim}, and more recently 
in \cite{blms,ms}. One argument in its favour is
that it seems hard to imagine how any notion of causality 
could emerge in the full quantum theory unless it had been
imposed in some form on the individual histories in the first place. 

Two-dimensional quantum gravity is an ideal testing ground 
for such ideas, since it can be solved explicitly.
We will show that by restricting the state sum to discrete 
geometries with a causal structure one obtains a theory of 
quantum gravity with the following features \cite{al,aal1}.
\begin{itemize}
\item[(1)] The expectation values of reparametrization-invariant
space-time distances have canonical 
dimensions. In other words, in spite of strong fluctuations in
the geometry, there is still a sense in which the quantum space-time 
is two-dimensional.
\item[(2)] When matter with conformal charge $c \leq 1$ is
coupled to this Lorentz\-ian quantum gravity theory, 
both property (1) and the properties of the conformal 
field theory (critical exponents etc.) remain unchanged.
\end{itemize}

There is an alternative way of arriving at the same theory,
namely, by starting from the geometric configurations of the
2d {\it Euclidean} gravity theory, and removing its ``baby universes''
in a systematic manner. (The proliferation of these branching
structures is responsible for most of the fractal and geometric 
properties of the Euclidean theory.) In this way one creates a many-to-1 
correspondence between Euclidean and Lorentzian geometries.
Lorentzian 2d quantum gravity appears 
in this construction as 
a ``renormalized'' version of the Euclidean quantum theory \cite{ackl}.
 
Loosening the requirement of causality for the individual 
space-time histories, one may allow for changes in the topology
of the {\it spatial} (constant-time) slices as a function of time, without 
changing the overall topology of {\it space-time}.  
This is of course tantamount to reintroducing geometries with
baby universes. 
We will show below that once this process is permitted,
it will totally dominate the structure of space-time. The resulting 
theory (which is equivalent to Euclidean Liouville quantum gravity) has 
the following features.
\begin{itemize}
\item[(1)] The expectation values of {\it geodesic} 
space-time distances have anomalous dimensions, and the intrinsic 
Hausdorff dimension of quantum space-time is four, and not two
\cite{kkmw,aw}.
\item[(2)] When matter with conformal charge $c \leq 1$ is
coupled to this Euclidean quantum gravity theory, both the fractal 
properties of the gravitational sector  
and the critical exponents of the conformal field theory are 
changed. 
\end{itemize}

We therefore may say that the
interaction between gravity and matter is weak 
in the Lorentzian 2d gravity model, at least as long as
$c\leq 1$. Allowing for baby-universe creation leads to a strong 
coupling between matter and gravity: the fractal properties of space-time
become a function of the matter content and in turn the back-reaction of 
the fluctuating geometry changes the critical properties of the matter.
The coupling of Ising models to 2d gravity provides a particularly
clear geometric illustration of the role played by the baby universes
\cite{baby}.

If one transcends the so-called $c=1$ barrier, even Lorentzian 
gravity exhibits a strong gravity-matter interaction,
leading to a change in the fractal properties of space-time
\cite{aal2}. In the Euclidean case (i.e. including baby universes),
it has been known for a long time that beyond $c=1$ the space-time 
disintegrates completely into fractals. These so-called branched 
polymers can be viewed as trees of baby universes of cut-off size. 
However, this is {\it not} what happens in the Lorentzian case.
One {\it does} observe a phase transition for $c > 1$ (to be precise,
the transition must take place somewhere in the interval
$1< c< 4$ \cite{aal2,aal3,aal4}), but the new geometry is less pathological 
than the branched polymers. In particular, the critical matter
exponents retain their Onsager values.

Many of the results quoted above for $c > 1$ have been obtained 
by Monte Carlo simulations.
Numerical methods have been very successful in the study
of discretized quantum gravity. 
Moreover, in two-dimensional quantum gravity there has been a 
fruitful interaction between theory and ``experiment''. Non-rigorous
calculations have been verified or falsified by numerical 
``experiments'', and ``observations'' have inspired theoretical 
progress. In the Lorentzian case, visualizations of the
computer-generated space-time geometries have been useful in
understanding the influence of matter on geometry \cite{website}. 

In higher dimensions, only partial results have been obtained
by analytical methods. These include asymptotic estimates for
the partition functions of dynamically triangulated 
Euclidean gravity in 3d and 4d \cite{carfora,acm}, 
and qualitative descriptions of the extreme branched-polymer and
crumpled phases \cite{acm,acm1,balls}.
However, as already mentioned, our regularized quantum gravity 
models can readily be studied by means of computer simulations.
In a first step, one determines the phase diagram of the discretized 
theory, in order to locate potential critical points where a 
continuum limit can be obtained. One can then use
standard finite-size tools to study the scaling of 
various observables. This method has 
been very successful and has enabled us to study both the fractal
properties of space-time, and the critical exponents 
of matter coupled to fluctuating geometries.

The rest of this paper is organized as follows. In Section \ref{model}
we describe and solve the simplest discretized Lorentzian model 
of 2d gravity, using a ``Lorentzian'' version of the method of
dynamical triangulations.
In Section \ref{continuum}, the corresponding continuum limit is
obtained, while in Section \ref{topology} we demonstrate how the 
inclusion of baby universes changes the structure of quantum space-time. 
A non-perturbative definition of 
Euclidean quantum gravity for arbitrary dimension
$d$ by means of dynamical 
triangulations is given in Section 
\ref{euclidean}, where we also discuss the inclusion of
matter fields. Section \ref{numerical} outlines  the 
principles for numerical simulations of the model, with special 
emphasis on 2d gravity. In Subsection \ref{observables}, we describe 
various possibilities of defining 
notions of (fractal) ``dimensionality''
in the framework of Euclidean quantum gravity.
We discuss why these ``observables'' are particularly well-suited 
for use in numerical calculations.
Subsection  \ref{results} provides the
interpretation of the numerical results obtained in 2d Euclidean 
quantum gravity. Section \ref{higherd} describes the numerical approach to 
higher-dimensional Euclidean quantum gravity. Finally, in Section 
\ref{outlook}, we outline the 
future perspectives for both Euclidean and Lorentzian 
lattice gravity.

\section{Lorentzian gravity in 2d}

\subsection{The discrete model}\label{model}

In order to solve two-dimensional quantum gravity in a 
path-integral formulation, one has to ``count'' geometries. 
Since for a fixed space-time topology the gravitational 
action consists only of the cosmological term, 
all geometries of a fixed space-time volume contribute with the 
same weight\footnote{Maybe surprisingly, in the 
framework of dynamical triangulations also higher-dimensional 
gravity can be reduced to a pure counting problem, c.f. 
Section \ref{funct}.}. In the simplest model of two-dimensional gravity,
no {\it spatial} topology changes are allowed either.
For simplicity, we choose the spatial slices (at constant ``time'')
to be circles, so that the overall topology 
of space-time is of the form $S^1\times [0,1]$ (for other 
spatial boundary conditions, see \cite{lotti}). 
This mimicks the situation in classical gravity, where 
one usually works with globally hyperbolic space-times \cite{waldbook}.
We will use a natural class of
triangulations of the cylinder $S^1\times [0,1]$, to which we will 
assign edge lengths (i.e. a discretized metric) in such a way
that each two-dimensional geometry carries a discrete causal structure.
Our simplicial space-times will have a preferred foliation into a 
discrete set of circular slices, consisting only of space-like
edges. The foliation parameter $t$ can be interpreted as a discrete
version of ``proper time''.
The main task of our dynamical-triangulations approach is to
count the geometries in this class, that is, to perform the state sum
in the regularized context, and then attempt to take a
continuum limit.

\begin{figure}
\centerline{\hbox{\psfig{figure=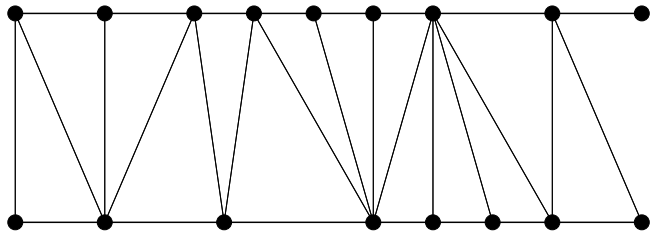,height=3cm,angle=0}}}
\caption[fig1]{The propagation of a spatial slice from step $t$ to 
step $t+1$. The ends of the strip should be joined to form a band
with topology $S^1 \times [0,1]$.}
\label{fig1}
\end{figure}

The geometry of each spatial slice is uniquely characterized by 
the length assigned to it. In the discretized version, the length $L$ 
will be ``quantized'' in units of a lattice spacing $a$, i.e.\ 
$L= l\cdot a$ where $l$ is an integer. A slice will thus be 
defined by $l$ vertices and $l$ links connecting them. 
To obtain a 2d geometry, we will evolve this spatial loop in 
discrete steps. This leads to a preferred notion of (discrete) ``time'' 
$t$, where each loop represents a slice of constant $t$.
The propagation from time-slice 
$t$ to time-slice $t+1$ is governed by the following rule: each vertex
$i$ at time $t$ is connected to $k_i$ vertices at time $t+1$, $k_i \geq 1$,
by links which are assigned squared edge lengths $-a^2$. 
The $k_i$ vertices, $k_i > 1$, 
at time-slice $t+1$ will be connected by $k_i-1$ consecutive 
space-like links, thus forming $k_i -1$ triangles. 
Finally the right boundary vertex
in the set of $k_i$ vertices will be identified with the left boundary 
vertex of the set of $k_{i+1}$ vertices. In this way we get a total of 
$\sum_{i=1}^l (k_i-1)$ vertices (and also links) at time-slice $t+1$ and 
the two spatial slices are connected by $\sum_{i=1}^l k_i
\equiv l_{t}+l_{t+1}$ triangles (see Fig.\ \ref{fig1}). 

The elementary building blocks of a geometry are therefore triangles
with one space- and two time-like edges. We define them to be flat
in the interior. A consistent way of assigning interior angles to
such Minkowskian triangles is described in \cite{sorkin}. The
angle between two time-like edges is $\gamma_{tt}=-\arccos \frac{3}{2}$,
and between a space- and a time-like edge $\gamma_{st}=
\frac{\pi}{2}+\frac{1}{2} \arccos \frac{3}{2}$, summing up to
$\gamma_{tt}+2\gamma_{st}=\pi$. The sum over all angles around a
vertex with $j$ incoming and $k$ outgoing time-like edges (by
definition $j,k\geq 1$) is given by $2\pi+(4-j-k)\arccos\frac{3}{2}$.
The regular triangulation of flat Minkowski space
corresponds to $j=k=2$ at all vertices. The volume of a single
triangle is given by $\frac{\sqrt{5}}{4}a^2$.

One may view these geometries as a subclass of all possible
triangulations that allow for the introduction of a causal
structure. Namely, if we think of all time-like links as being
future-directed, a vertex $v'$ lies in the future of a
vertex $v$ iff there is an oriented sequence of time-like
links leading from $v$ to $v'$. Two arbitrary vertices may or may not
be causally related in this way. 

In quantum gravity we are instructed to sum over all geometries connecting, 
say, two spatial boundaries of length $L_1$ and $L_2$, with the weight 
of each geometry $g$ given by 
\beq\label{1}
\e^{i S[g]}, ~~~~~S[g] = \L \int \!\!\sqrt{-\det g}~~~(\mbox{in 2d}),
\eeq
where $\L$ is the cosmological constant.
In our discretized model the boundaries will be characterized by 
integers $l_1$ and $l_2$, the number of vertices or links at the two
boundaries. The path-integral amplitude for the propagation from 
geometry $l_1$ to $l_2$ will be the sum over all interpolating 
surfaces of the 
kind described above, with a weight given by the discretized version of 
\rf{1}. Let us call the corresponding amplitude $G^{(1)}_\l(l_1,l_2)$.
We thus have
\bea
G_\l^{(1)}(l_1,l_2) &=& \sum_{t=1}^{\infty} G_\l^{(1)}(l_1,l_2;t),\label{3}\\
G_\l^{(1)}(l_1,l_2;t) &=& 
\sum_{l=1}^\infty G_\l^{(1)}(l_1,l;1)\;l\;G_\l^{(1)}(l,l_2,t-1),\label{4}\\
G_\l^{(1)}(l_1,l_2;1) &=& \frac{1}{l_1}\sum_{\{k_1,\dots,k_{l_1}\}} 
\e^{i \l a^2 \sum_{i=1}^{l_1} k_i}, \label{5}
\eea
where $\l$ denotes the {\em bare}  
cosmological constant\footnote{One obtains 
the renormalized (continuum) cosmological constant $\L$ in \rf{1} by 
an additive renormalization, see below.} (we have absorbed the finite
triangle volume factor), and where $t$ denotes the 
total number of time-slices connecting $l_1$ and $l_2$. 

From a combinatorial point of view it is convenient to mark a 
vertex on the entrance loop in order to get rid of the factors
$l$ and $1/l$ in \rf{4} and \rf{5}, that is,
\beq\label{6}
G_\l (l_1,l_2;t) \equiv l_1 G_\l^{(1)}(l_1,l_2;t)
\eeq
(the unmarking of a point may be thought of as 
the factoring out by (discrete) spatial diffeomorphisms).
Note that $G_\l(l_1,l_2;1)$ plays the role of a 
transfer matrix, satisfying
\bea
G_\l(l_1,l_2;t_1+t_2) &=& \sum_{l} G_\l(l_1,l;t_1)\; G_\l(l,l_2;t_2)\label{7}\\
G_\l(l_1,l_2;t+1) &=& \sum_{l} G_{\l}(l_1,l;1)\;G_\l(l,l_2;t).\label{8}
\eea
Knowing $G_\l(l_1,l_2;1)$ allows us to find $G_\l(l_1,l_2;t)$
by iterating \rf{8} $t$ times. This program is conveniently 
carried out by introducing the generating function for the numbers
$G_\l(l_1,l_2;t)$,
\beq\label{9}
G_\l(x,y;t)\equiv \sum_{k,l} x^k\,y^l \;G_\l(k,l;t),
\eeq
which we can use to rewrite \rf{7} as 
\beq\label{10}
G_\l(x,y;t_1+t_2) = \ointz G_\l(x,z^{-1};t_1) G_\l(z,y;t_2),
\eeq
where the contour should be chosen to include the singularities 
in the complex $z$--plane of $G_\l(x,z^{-1};t_1)$ but not those
of $G_\l(z,y;t_2)$. 

One can either view the introduction of $G_\l(x,y;t)$ as a purely
technical device or take $x$ and $y$ as boundary cosmological 
constants,
\beq\label{10a}
x=\e^{i\l_ia},~~~~y=\e^{i\l_oa},
\eeq
such that $x^k= \e^{i\l_i a\,k}$ becomes a boundary cosmological term,
and similarly for $y^l= \e^{i\l_o a\, l}$. 
Let us for notational convenience define 
\beq\label{11}
g=\e^{i\l a^2}
\eeq
(not to be confused with the symbol for the continuum metric).
For the technical purpose of counting we view $x,y$ and $g$ as 
variables in the complex plane. In general the function 
\beq\label{11a}
G(x,y;g;t)\equiv G_\l(x,y;t)
\eeq
will be analytic in a neighbourhood of $(x,y,g)=(0,0,0)$.  

From the definitions \rf{5} and \rf{6} it follows by standard techniques 
of generating functions that we may associate a factor $g$ with each 
triangle, a factor $x$ with each vertex on the entrance loop and 
a factor $y$ with each vertex on the exit loop, leading to
\beq\label{12}
G(x,y;g;1) =\sum_{k=0}^\infty \left( gx \sum_{l=0}^\infty
 (gy)^l \right)^k -
\sum_{k=0}^\infty (gx)^k = \frac{g^2 xy}{(1-gx)(1-gx-gy)}.
\eeq
Formula \rf{12} is simply a book-keeping device for all possible
ways of evolving from an entrance loop of any length in one step to
an exit loop of any length. The subtraction of the term $1/(1-gx)$ 
has been performed to 
exclude the degenerate cases where either the entrance or the exit 
loop is of length zero. 
  
From \rf{12} and eq.\ \rf{10}, with $t_1=1$, we obtain
\beq\label{13}
G(x,y;g;t) = \frac{gx}{1-gx}\; G(\frac{g}{1-gx},y;g;t-1).
\eeq
This equation can be iterated and the solution written as 
\beq\label{14}
G(x,y;g;t) = F_1^2(x)F_2^2(x) \cdots F_{t-1}^2(x) 
\frac{g^2 xy}{[1-gF_{t-1}(x)][1-gF_{t-1}(x)-gy]},
\eeq
where $F_t(x)$ is defined iteratively by
\beq\label{15}
F_t(x) = \frac{g}{1-gF_{t-1}(x)},~~~F_0(x)=x.
\eeq
Let $F$ denote the fixed point of this iterative equation. By standard
techniques one readily obtains
\beq\label{16}
F_t(x)= F\ \frac{1-xF +F^{2t-1}(x-F)}{1-xF +F^{2t+1}(x-F)},~~~~
F=\frac{1-\sqrt{1-4g^2}}{2g}.
\eeq
Inserting \rf{16} in eq.\ \rf{14}, we can write
\bea
&&G(x,y;g,t) = \frac{ F^{2t}(1-F^2)^2\; xy}
{(A_t-B_tx)(A_t-B_t(x+y)+C_txy)}
\label{17}\\
&&\hspace{.3cm}  =
 \frac{F^{2t}(1-F^2)^2\;xy}{\Big[(1\!\!-\!xF)\!-\!F^{2t+1}(F\!\!-\!x)\Big]
\Big[(1\!\!-\!xF)(1\!\!-\!yF)\!-\!F^{2t} (F\!\!-\!x)(F\!\!-\!y)\Big]},
~~~\label{17a}
\eea
where the time-dependent coefficients are given by 
\beq\label{18}
A_t =1-F^{2t+2},~~~B_t=F(1-F^{2t}),~~~C_t=F^2(1-F^{2t-2}).
\eeq
The combined region of convergence to the 
expansion in powers $g^kx^ly^m$, valid for all $t$ is 
\beq\label{18a}
|g| < \oh,~~~~ |x|< 1,~~~~|y|<1.
\eeq

\subsection{The continuum limit}\label{continuum}

The path integral formalism we are using here
is very similar to the one used to re\-pre\-sent the free particle as 
a sum over paths. Also there one performs a
summation over geometric objects (the paths), and the path integral itself
serves as the propagator. From the particle case it is known that the bare mass
undergoes an additive renormalization (even for the free particle), 
and that the bare propagator is subject to a wave-function renormalization
(see \cite{book} for a review). The same is true
in two-dimensional Euclidean gravity, treated in the formalism of 
dynamical triangulations \cite{book}. The coupling constants
with positive mass dimension, i.e.\ the cosmological constant and the 
boundary cosmological constants, undergo an 
additive renormalization, while the partition function itself (i.e.\ the 
Hartle-Hawking-like wave function) 
undergoes a multiplicative wave-function renormalization.
We therefore expect the bare coupling constants $\l,\l_i$ and 
$\l_0$ to behave as 
\beq\label{20a}
\l = \frac{C_{\l}}{a^2} + \tilde{\L},~~~~
\l_i= \frac{C_{\l_{i}}}{a}+\tilde{X},~~~
\l_o =\frac{C_{\l_{o}}}{a}+\tilde{Y},
\eeq
where $\tilde{\L},\tilde{X},\tilde{Y}$ denote the renormalized 
cosmological and boundary cosmological constants. If we introduce
the notation 
\beq\label{20c}
g_c = \e^{i C_{\l}},~~~~x_c= \e^{i C_{\l_{i}}},~~~~y_c=\e^{iC_{\l_{o}}},
\eeq
for critical values of the coupling constants, 
it follows from \rf{10a} and \rf{11} that 
\beq\label{20b}
g=g_c\,\e^{ia^2\tL},~~~~x=x_c\,\e^{ia\tX},~~~~y=y_c\,\e^{ia\tY}.
\eeq 
The wave-function renormalization will appear as a multiplicative  
cut-off dependent factor in front of the bare 
``Green's function'' $G(x,y;g;t)$, 
\beq\label{20}
G_\tL (\tX,\tY;T) = \lim_{a \to 0} a^{\eta} G(x,y;g;t),
\eeq
where $T=a\, t$, and where the critical exponent $\eta$ 
should be chosen so that 
the right-hand side of eq.\ \rf{20} exists. In general this will only be 
possible for particular
choices of $g_c,x_c$ and $y_c$ in \rf{20}. 

The basic relation \rf{7} can survive the limit \rf{20} only 
if $\eta=1$, since we have assumed that the boundary lengths 
$L_1$ and $L_2$ have canonical dimensions and satisfy $L_i = a\, l_i$.

A closer analysis reveals that only at $g_c=1/2$
one can obtain a sensible continuum limit.
 It corresponds to 
a purely imaginary bare cosmological constant
$\l_{c}:=C_{\l}/a^{2} = -i \ln 2/a^2$.
If we want to approach this point from the region in the 
complex $g$-plane where $G(x,y;g;t)$  
converges it is natural to choose the renormalized coupling $\tL$
imaginary as well, $\tL = i\L$, i.e.
\beq\label{25a}
\l = i\ \frac{\ln \oh}{a^2} +i \L.
\eeq
One obtains a well-defined scaling limit (corresponding to
$\L\in \R$) by letting $\l\to \l_{c}$ along the imaginary axis.
The Lorentzian form for the continuum 
propagator is obtained by an analytic continuation $\L\to -i\L$
in the {\it renormalized} coupling of the resulting Euclidean 
expressions. 

At this stage it may seem that we are surreptitiously reverting
to a fully Euclidean model. We could of course equivalently 
have conducted
the entire discussion up to this point in the ``Euclidean sector'',
by omitting the factor of $-i$ in the exponential \rf{1} of the
action, choosing $\l$ positive real and taking all edge lengths 
equal to 1. However, from a purely Euclidean point of view there
would not have been any reason for restricting the state sum to a subclass
of geometries admitting a causal structure. The associated 
preferred notion of a discrete ``time'' allows us to define
an ``analytic continuation in time''. Because of the simple
form of the action in two dimensions, the rotation 
\beq\label{25b}
\int  dx\ dt \sqrt{-\det g_{lor}} \to i\int  dx \ dt_{eu} 
\sqrt{\det g_{eu}}
\eeq
to Euclidean metrics in our model is equivalent to the analytic continuation
of the cosmological constant $\L$.

From \rf{17} or \rf{17a} it 
follows that we can only get macroscopic loops in the limit 
$a \to 0$ if we simultaneously take $x,y \to 1$. (For $g_c=-1/2$, one
needs to take $x,y \to -1$. The continuum expressions one obtains
are identical to those for $g_c=1/2$.) Again 
the critical points correspond to purely imaginary 
bare boundary cosmological coupling constants. We will 
allow for such imaginary couplings and thus approach the 
critical point $\l_i= \l_o=0$ from the region of convergence of 
$G(x,y;g;t)$, i.e.\ via real, positive $X,Y$ where  
\beq\label{25c}
\l_i = i X a,~~~~\l_o=i Y a.
\eeq
Again $X$ and $Y$ have an obvious interpretation as positive boundary
cosmological constants in a Euclidean theory, which may be
analytically continued to imaginary values to reach the Lorentzian
sector.

Summarizing, we have 
\beq\label{25}
g=\oh \e^{-\L a^2} \to \oh (1-\L a^2),~~~(\mbox{i.e.}~~F=1-a\sqrt{\L})
\eeq
as well as 
\beq\label{25d}
x=\e^{-Xa} \to 1-aX,~~~~~~y=\e^{-aY} \to 1-aY,
\eeq
where the arrows $\to$ in \rf{25} and \rf{25d} should be viewed 
as analytic coupling constant redefinitions of $\L,X$ and $Y$,
which we have performed to get rid of factors of 1/2 etc. in the formulas
below.
With the definitions \rf{25} and \rf{25d} it is straightforward 
to perform the continuum limit of $G(x,y;g;t)$ as $(x,y,g) \to 
(x_c,y_c,g_c)=(1,1,1/2)$, yielding
\bea
&&G_\L(X,Y;T) = \frac{4\L\ \e^{-2\SLT}}{(\SL+X)+\e^{-2\SLT}(\SL-X)}\nonumber\\
&&\hspace{.5cm}\times \, \frac{1}{(\SL+X)(\SL+Y)-\e^{-2\SLT}(\SL-X)(\SL-Y)}.
\label{26}
\eea
For $T \to \infty$ one finds 
\beq\label{27}
G_\L(X,Y;T) \buildrel{T\rightarrow
\infty}\over\longrightarrow \frac{4\L \;\e^{-2\SLT}}{(X+\SL)^2(Y+\SL)},
\eeq
while the limit for $T \to 0$ is
\beq\label{28}
G_\L(X,Y;T\equ 0) = \frac{1}{X+Y},
\eeq
in accordance with the expectation that 
\beq\label{29}
G_\L(L_1,L_2;T\equ 0) = \del(L_1-L_2).
\eeq

The general expression for $G_\L(L_1,L_2;T)$ can be computed
as the inverse Laplace transform 
of formula \rf{26}, yielding
\beq\label{30}
G_\L(L_1,L_2;T) = \frac{\e^{-[\coth \SLT] \SL(L_1+L_2)}}{\sinh \SLT}
\; \frac{\sqrt{\L L_1 L_2}}{L_2}\; \; 
I_1\left(\frac{2\sqrt{\L L_1 L_2}}{\sinh \SLT}\right), 
\eeq
where $I_1(x)$ is a modified Bessel function of the first kind.
The asymmetry between $L_1$ and $L_2$ arises because the entrance loop 
has a marked point, whereas the exit loop has not. The amplitude with 
both loops marked is obtained by multiplying with $L_2$, while the 
amplitude with no marked loops is obtained after dividing 
\rf{30} by $L_1$. Quite remarkably, our highly non-trivial 
expression \rf{30} agrees
with the loop propagator obtained from a bona-fide continuum calculation
in proper-time gauge of pure 2d gravity by Nakayama \cite{nakayama}.

The basic result \rf{26} for $G_\L(X,Y;T)$
can be derived by taking the continuum limit of 
the recursion relation  \rf{13}. By inserting \rf{25} and \rf{25d} 
in eq.\ \rf{13}
and expanding to first order in the lattice spacing $a$ we obtain
\beq\label{32} 
\frac{\prt}{\prt T} G_\L(X,Y;T) + \frac{\prt}{\prt X}
\Bigl[ (X^2-\L) G_\L(X,Y;T) \Bigr]=0.
\eeq
This is a standard first-order partial differential equation which 
should be solved with the boundary condition \rf{28} at $T=0$, since this
expresses the natural condition \rf{29} on $G_\L(L_1,L_2)$.
The solution is thus 
\beq\label{33}
G_\L(X,Y;T) = \frac{\bar{X}^2(T;X)-\L}{X^2-\L}\; \frac{1}{\bar{X}(T;X)+Y}, 
\eeq
where $\bar{X}(T;X)$ is the solution to the characteristic equation
\beq\label{34}
\frac{d \bar{X}}{dT} = -(\bar{X}^2-\L),~~~~\bar{X}(T=0)=X.
\eeq
It is readily seen that the solution is indeed given by \rf{26}
since we obtain
\beq\label{35}
\bar{X}(T) = \SL \; 
\frac{(\SL+X)-\e^{-2\SLT}(\SL-X)}{(\SL+X)+\e^{-2\SLT}(\SL-X)}.
\eeq

If we interpret the propagator $G_\L(L_1,L_2;T)$ as the matrix element
between two boundary states of a Hamiltonian evolution in 
``time'' $T$,
\beq\label{ham}
G_\L(L_1,L_2;T)=\langle L_1|\e^{-\hat H T}|L_2\rangle
\eeq 
we can, after an inverse Laplace transformation, read off the functional form
of the Hamiltonian operator $\hat H$ from \rf{32},
\beq\label{35b}
\hat H(L,\frac{\partial}{\partial L})=
 -L\frac{\partial^2}{\partial L^2}+L\L .
\eeq
 
The corresponding Hamiltonian for the propagator of unmarked loops
is given by
\beq\label{35e}
\hat H_u(L,\frac{\partial}{\partial L})=
-L \frac{\partial^2}{\partial L^2}-2 \frac{\partial}{\partial L}
+\L L.
\eeq

The Hamiltonian \rf{35e} is Hermitian with respect to the natural
measure $L\,\d L$, which has its origin in the  
basic completeness relation \rf{4} for the transfer matrix
with unmarked entrance and exit loops.
If one wants to construct a unitary evolution with 
respect to the ``time''-parameter $T$ appearing in the
transfer-matrix approach, one can simply exponentiate
$(i\hat H T)$. 

However, we should point out that there is an alternative to the 
analytic continuation $T \to -iT$ if one wants to relate 
the Euclidean and Lorentzian sectors of the theory. 
The combination $\SL T$ appearing as an argument in \rf{30}
arises in taking the continuum limit of powers of the form $F^{t}$
in expressions like \rf{17}, \rf{17a}, where $F$ is defined in \rf{16}.
There are two aspects to a possible analytic continuation of 
$F^t$. The power $t$ in $F^t$ should clearly not be continued,
since it is simply an integer counting
the number of iterations of the transfer matrix. 
However, the function $F$ itself does refer
to the action, because the dimensionless coupling constant
$g = \e^{i\l a_t a_l}$ is the action for a single Lorentzian triangle. 
(For added clarity we have distinguished between the lattice spacings
in time- and space-directions, and called them $a_{t}$ and $a_{l}$.)
From the expression for $F$ in terms of $g$ in \rf{16}, we have
$F=1-\sqrt{a_{t}a_{l}\L}$.
The analytic continuation of $F$ in time, from Euclidean to Lorentzian 
time, corresponds to the substitution $a_{t}\to -i\ a_{t}$ 
under the square-root sign, and thus becomes equivalent 
to the continuation $\L \to -i\L$ in the cosmological constant, 
as already remarked below eq.\ \rf{25b}. 
The subtleties associated with the 
analytic continuation in the ``time''-parameter $T$ appearing in a 
transfer-matrix formulation of quantum gravity 
were first discussed in \cite{jeff1,jeff2} in the context of a
square-root action formulation.
Similar difficulties will also be present in higher-dimensional
gravity, where the analytic continuation from Euclidean metrics to 
Lorentzian metrics cannot be absorbed by
a continuation in $\L$ alone. To conclude, it 
is not obvious how to choose the correct analytic continuation 
back to Lorentzian signature, once the continuum limit has been taken.
The continuation $T \to -iT$ leads to a unitary theory. 
For the continuation  $\L \to -i\L$, we have not yet found a
scalar product which makes the corresponding
evolution operator unitary.

\section{Topology changes and Euclidean quantum gravity}\label{topology}

\subsection{Baby universe creation}\label{baby}

In our non-perturbative regularization of 2d quantum gravity we have
so far not included the possibility of topology changes of space.
We will now show that 
{\em if} one allows for spatial topology changes, one is led in
an essentially unambiguous manner to a {\it different} 
theory  of two-dimensional quantum gravity, where space-time is 
much more fractal, and which agrees with Euclidean 2d quantum 
gravity as defined by Euclidean 
dynamical triangulations or Liouville theory.

By a topology change of {\em space} in our Lorentzian setting
we have in mind the following: a baby universe may
branch off at some time $T$ and develop in the future, where it 
will eventually disappear in the vacuum, but it is
not allowed to rejoin the ``parent'' universe and thus change the 
overall topology of the two-dimensional manifold. This is 
a restriction we impose to be able to compare with 
the analogous calculation in usual 2d Euclidean quantum gravity.

It is well known that such a branching leads to additional
complications, compared with the Euclidean situation, in the sense
that, in general, no continuum Lorentzian metrics which are smooth and 
non-degenerate everywhere can be defined on such space-times (see, for
example, \cite{ls} and references therein). 
These considerations do not affect the cosmological term in the
action, but lead potentially to contributions from the Einstein-Hilbert
term at the singular points where a branching or pinching occurs.

We have so far ignored the curvature term in the action since it
gives merely a constant contribution in the absence of topology
change. We will continue to do so in the slightly generalized
setting just introduced. The continuum results of \cite{ls} 
suggest that the contributions from the two singular points 
associated with each branching of a baby universe (one at the
branching point and one at the tip of the baby universe where it
contracts to a point) cancel in the action.
The physical geometry of these configurations may 
seem slightly contrived, but they may well be 
important in the quantum theory of gravity and deserve further 
study. However, for the moment our main motivation for introducing them
is to make contact with the 
usual non-perturbative Euclidean path-integral results.

We will use the rest of this section to demonstrate the following:
once we allow for spatial topology changes, 
\begin{itemize}
\item[(1)]this process completely dominates and changes the critical 
behaviour of the discretized theory, and
\item[(2)] the disc amplitude $W_\L(L)$ (the Hartle-Hawking wave function)
is uni\-que\-ly determined, almost without any calculations.
\end{itemize}
Our starting point will be the discretized model introduced in
Sec.\ \ref{model}. In this model we do not directly have a 
disc amplitude like the Hartle-Hawking wave functional. However,
as discussed at the beginning of this section, the degeneracy of the 
metric at a point in the interior of the disc is always compensated 
(in the sense of complex contributions to the action) by the degeneracy of the
metric at the point where the baby universe branched off. We can thus define 
the disc amplitude in Lorentzian gravity as 
\beq\label{topz1}
w^{(b)}(x,g) := \sum_t G^{(b)}(x,l_2\equ 1;g;t) =G^{(b)}(x,l_2\equ 1;g),
\eeq
and the continuum version as
\beq\label{topz1a}
W^{(b)}_\L (X) = \int \d T \; G^{(b)}_\L (X,L_2\equ 0;T)  = 
G^{(b)}_\L (X,L_2\equ 0)
\eeq
where the superscript $^{(b)}$ indicates the 
``bare'' Lorentzian model without spatial topology changes.
It follows that  
\beq\label{topz2}
w^{(b)}(x,g) \to a^{-1} W^{(b)}_\L (X).
\eeq

\begin{figure}
\centerline{\hbox{\psfig{figure=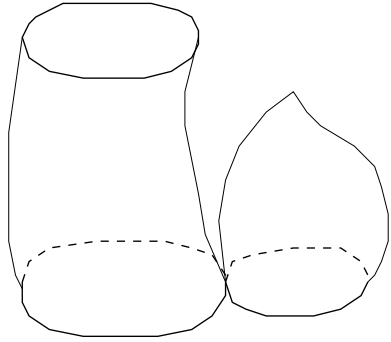,height=4.5cm,angle=0}}}
\caption[topchange]{A ``baby universe''  created by a global pinching.}
\label{topchange}
\end{figure}

There are a number of ways to implement the creation of baby universes, 
some more natural than others, but they all agree in the continuum limit,
as will be clear from the general arguments provided below. 
Here we discuss only the simplest way of implementing such a change.
This is shown in Fig.\ \ref{topchange}: stepping forward from 
$t$ to $t+1$ from a loop of length $l_1$ we create a
baby universe of length $l < l_1$ by pinching it off non-locally
from the main branch. 

Accounting for the new possibilities of evolution in each step
according to fig.\ \ref{topchange}, the new and old transfer matrices
are related by 
\beq\label{top1}
G_\l(l_1,l_2;1) = G_\l^{(b)}(l_1,l_2;1)+ 
\sum_{l=1}^{l_1-1} l_1 w(l_1\mi l,g)\,G_\l^{(b)}(l,l_2;1).
\eeq
The factor $l_1$ in the sum comes from the fact that the 
``pinching'' shown in fig.\ \ref{topchange} can take place at any of the 
$l_1$ vertices. As before, the new transfer matrix leads to new amplitudes
$G_\l(l_1,l_2;t)$, satisfying
\beq\label{top2a}
G_\l(l_1,l_2;t_1+t_2) = \sum_l G_\l(l_1,l;t_1)G_\l(l,l_2;t_2),
\eeq
and in particular 
\beq\label{top2}
G_\l(l_1,l_2;t) = \sum_l G_\l(l_1,l;1)\;G_\l(l,l_2;t\mi 1).
\eeq
Performing a (discrete) Laplace transformation of eq.\ \rf{top2}
leads to
\bea
{G(x,y;g;t) =}~~~~~~~~~~~~~~~~~~~~~~~~~~~~~~~~~~~~~~~~~~~~~~~~~~~~~~~
~~~~~~~~~~~~~~~~~~~~~~
 & &  \nonumber \\ 
\ointz  \left[G_\l^{(b)}(x,z^{-1};1)\pl x \frac{\prt}{\prt x} 
\Bigl( w(x;g) G_\l^{(b)}(x,z^{-1};1)\Bigr) \right]  G(z,y;g;t \mi 1),
&&\nonumber \\
\!\!\!\!\!\!\!\! \label{top3}
\eea
or, using the explicit form of the transfer matrix $G_\l^{(b)}(x,z;1)$,
formula \rf{12},
\beq\label{top4}
G(x,y;g;t) = \Bigl[1+x\frac{\prt w(x,g)}{\prt x}
+ x w(x,g)\frac{\prt}{\prt x} \Bigr]
\, \frac{gx}{1-gx} \, G\Bigl( \frac{g}{1\mi gx},y;g;t\mi 1\Bigr).
\eeq
At this point neither the disc amplitude $w(x,g)$ 
nor $G(x,y;g;t)$ are known. We will now 
show that they are uniquely determined if we assume 
that the boundary length scales canonically with the lattice
spacing, $L=a \, l$, implying a renormalized boundary 
cosmological constant $X$ with the dimension of mass, $x = x_c(1-aX)$.
In addition we assume that 
the dimension of the renormalized cosmological constant 
$\L$ is canonical too, $g=g_c(1-\oh \L a^2)$. Somewhat related 
arguments have been presented in different settings in \cite{ik,watabiki}.

It follows from relation \rf{top2a} that we need 
\beq\label{top7}
G_\l(l_1,l_2,t) \buildrel{a\rightarrow
0}\over\longrightarrow  a\, G_\L(L_1,L_2;T).
\eeq 
It is important for the following discussion that $G_\l(l_1,l_2;t)$
cannot contain a non-scaling part since from first principles (sub-additivity) 
it has to decay exponentially in $t$.
By a Laplace transformation, using $x=x_c(1-a X)$ in the 
scaling limit, we thus conclude that 
\beq\label{top8}
G_\l(x,l_2,t) \buildrel{a\rightarrow
0}\over\longrightarrow G_\L(X,L_2,T),
\eeq
and further, by a Laplace transformation in $L_2$,  
\beq\label{top8x}
G_\l(x,y;t) \buildrel{a\rightarrow
0}\over\longrightarrow a^{-1} G_\L(X,Y;T).
\eeq

We will now show that the scaling of $w(x,g)$ is quite restricted 
too. The starting point is a combinatorial identity which the 
disc amplitude has to satisfy. The arguments are valid both 
for the disc amplitude in Euclidean quantum gravity and the 
disc amplitude we have introduced for our model in \rf{topz1}. 
We will assume the general scaling 
\beq\label{an3}
w(x,g)= w_{ns}(x,g)+ a^{\eta}W_\L(X) + \mbox{less singular terms}
\eeq
for the disc amplitude. In the case $\eta < 0$ the first term is 
considered absent (or irrelevant). 
However, if $\eta >0$  a term like $w_{ns}$ will generically 
be present, since any slight redefinition of the coupling constants of the 
model will produce such a term if it was not there from the beginning.

\begin{figure}
\centerline{\hbox{\psfig{figure=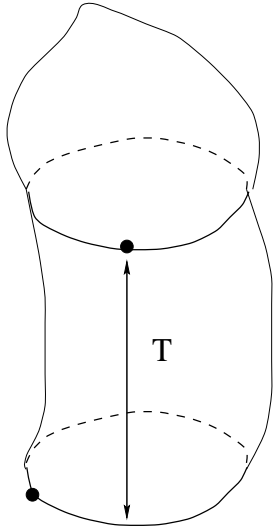,height=7cm,angle=0}}}
\caption[identity]{Marking a vertex in the bulk of $W_\L(X)$. The mark
has a distance $T$ from the boundary loop, which itself has one marked vertex.}
\label{identity}
\end{figure}

We will introduce an explicit mark in the bulk of $w(x,g)$
by differentiating with respect to $g$. This leads to the 
combinatorial identity 
\beq\label{an1}
g\ \frac{\prt w(x,g)}{\prt g} =
\sum_t \sum_l G(x,l;g;t) \, l\, w(l,g),
\eeq
or, after the usual Laplace transform,
\beq\label{an2}
g\ \frac{\prt w(x,g)}{\prt g}=
\sum_t \ointz G(x,z^{-1};g;t)\; \frac{\prt w(z,g)}{\prt z}.
\eeq 
The situation is illustrated in Fig.\ \ref{identity}. 
A given mark has a 
distance $t$ ($T$ in the continuum) to the entrance loop. 
In the figure we have drawn all points which have the same distance 
to the entrance loop and which form a connected loop. In the 
bare model these are all the points at distance $t$. 
In the case where baby universes are allowed (which we have not
included in the figure),
there can be many disconnected loops at the same distance.
Let us assume a general scaling 
\beq\label{an3a}
T = a^{\ep} t,~~~~~\ep >0,
\eeq
for the time variable $T$ in the continuum limit. Above we saw that 
the bare model without baby universe creation corresponded to $\ep=1$. 
With the generalization \rf{an3a} we account for the fact that
by allowing for baby universes
we have introduced an explicit asymmetry between 
the time- and space-directions. 

Inserting \rf{an3} and \rf{an3a} into eq.\ \rf{an2} we obtain
\bea
&&\frac{\prt w_{ns}}{\prt g}- 2a^{\eta-2} 
\frac{\prt W_\L(X)}{\prt \L} = \nonumber \\
&&\hspace{.6cm} \frac{1}{a^{\ep}} 
  \int \d T \int dZ\;  G_\L(X,-Z;T) 
\bigg[ \frac{\prt w_{ns}}{\prt z} -a^{\eta-1}
\frac{1}{z_{c}} \frac{\prt W_\L(Z)}{\prt Z}
\bigg],
\label{an4}
\eea
where $(x,g)=(x_c,g_c)$ in the non-singular part. 

From eq.\ \rf{an4} and the requirement $\epsilon >0$ it follows that the 
only consistent choices for $\eta$ are
\begin{itemize}
\item[{\bf 1:}] {\boldmath{${\eta <\ }$}{\bf 0}},~~ i.e.\
\beq\label{an5c}
a^{\eta-2} \frac{\prt W_\L(X)}{\prt \L} = \frac{a^{\eta-1}}{2 a^{\ep}} 
  \int \d T \int dZ\;  G_\L(X,-Z;T) \;\frac{1}{z_{c}}
  \frac{\prt W_\L(Z)}{\prt Z},
\eeq
in which case we get $\ep =1$; and
\item[{\bf 2:}] {\bf 1}{\boldmath${\ < \eta < \ }$}{\bf 2}. 
Here formula \rf{an4} splits into the two 
equations 
\beq\label{an5}
-a^{\eta-2} \frac{\prt W_\L(X)}{\prt \L} = \frac{1}{2 a^{\ep}}\, 
\frac{\prt w_{ns}}{\prt z}\bigg|_{z=x_c}\;
  \int \d T \int dZ\;  G_\L(X,-Z;T), 
\eeq  
and 
\beq\label{an6}
\frac{\prt w_{ns}}{\prt g}\bigg|_{g=g_c} = -\frac{a^{\eta-1}}{a^{\ep}}  
  \int \d T \int dZ\;  G_\L(X,-Z;T) 
\;\frac{1}{z_{c}} \frac{\prt W_\L(Z)}{\prt Z}.
\eeq
We are led to the conclusion that $\ep= 1/2$ and $\eta=3/2$, which are
precisely the values found in Euclidean 2d gravity.
Let us further remark that eq.\ \rf{an5} in this case becomes
\beq\label{an5a}
-\frac{\prt W_\L(X)}{\prt \L} = \mbox{const.}\;G_\L(X,L_2=0),
\eeq
which differs from \rf{topz1a} by a derivative with respect to the 
cosmological constant. We will explain the reason for this 
difference below. 
Finally, eq.\ \rf{an6} becomes
\beq\label{an7}
\int \d T \int dZ\;  G_\L(X,-Z;T) 
\;\frac{\prt W_\L(Z)}{\prt Z} = \mbox{const.},
\eeq
which will be satisfied automatically if $\eta =3/2$ and $\ep=1/2$,
as we will show below.  
\end{itemize}

We will now analyze a possible scaling limit of \rf{top4}, 
assuming the canonical scaling 
$x=x_c(1-aX)$ and $g=g_c(1-\oh \L a^2)$.
In order that the equation have a scaling limit at all, $x_c,\, g_c$ and
$w_{ns}(x_c,g_c)$ must satisfy two relations which can be determined
straightforwardly from \rf{top4}. The remaining continuum equation reads
\bea
a^\ep\frac{\prt}{\prt T}\, G_\L(X,Y;T)& =&
-a \, \frac{\prt}{\prt X} \Bigl[ (X^2-\L) G_\L(X,Y;T)\Bigr] \nn
&&-a^{\eta-1}\frac{\prt}{\prt X} \Bigl[W_\L(X) G_\L(X,Y;T)\Bigr].
\label{top13}
\eea
The first term on the right-hand side of eq.\ \rf{top13} 
is precisely the one we have already encountered 
in our original model, while the second term 
is due to the creation of baby universes. 
Clearly the case $\eta < 0$ (in fact $\eta \leq 1$) is inconsistent
with the presence of the second term, i.e.\ the creation of baby universes. 
However, since $\eta <2$, 
the last term on the right-hand side of \rf{top13} will always 
dominate over the first term. {\em Once we 
allow for the creation of baby universes, this process will completely
dominate the continuum limit.} In addition we get $\ep = \eta-1$,
in agreement with \rf{an6}. It follows that $\eta > 1$ and 
we conclude that $\ep=1/2,\eta=3/2$ 
are the only possible scaling exponents if we allow for the creation 
of baby universes. 
These are precisely the scaling exponents obtained from 
two-dimensional Euclidean gravity in
terms of dynamical triangulations, as we have already remarked. The topology 
changes of space have induced an anomalous dimension for $T$. 
If the second term on the right-hand side of \rf{top13} had been absent,
this would have led to $\ep =1$, and the time $T$ scaling in the 
same way as the spatial length $L$.

In summary, in the case $(\eta,\ep)=(3/2,1/2)$ eq.\ \rf{top13} leads 
to the continuum equation
\beq\label{top16}
\frac{\prt}{\prt T}\, G_\L(X,Y;T) =
-\frac{\prt}{\prt X} \Bigl[W_\L(X) G_\L(X,Y;T)\Bigr],
\eeq 
which, combined with eq.\ \rf{an5a}, determines the continuum disc
amplitude $W_\L(X)$.
Integrating \rf{top16} with respect to $T$ and using that 
$G_\L(L_1,L_2;T\equ 0)=\delta(L_1\mi L_2)$, i.e.\
\beq\label{top18a} 
G_\L(X,L_2\equ 0;T\equ 0)=1,
\eeq
we obtain
\beq\label{top18}
-1 = \frac{\prt}{\prt X}\bigg[ W_\L(X) \frac{\prt}{\prt\L} W_\L(X)
\bigg].
\eeq
Since $W_\L(X)$ has length dimension --3/2, i.e. 
$W_\L^2(X) = X^3 F(\SL/X)$, the general solution must be of the form 
\beq\label{top19}
W_\L(X) = \sqrt{-2\L X + b^2 X^3+ c^2 \L^{3/2}}.
\eeq
From the very origin of $W_\L(X)$ as the Laplace transform of a disc 
amplitude $W_\L(L)$ which is bounded, it follows that $W_\L(X)$ has 
no singularities or cuts for $\mbox{Re}\, X >0$. This requirement 
fixes the constants $b,c$ in \rf{top19} such that 
\beq\label{top20}
W_\L(X) = b \Big(X-\frac{\sqrt{2}}{b\,\sqrt{3}} \,\SL\Big)
\sqrt{X+ \frac{2\sqrt{2}}{b\,\sqrt{3}}\SL},
\eeq
where the constant $b$ is determined by the model-dependent 
constant in \rf{an5a}. 
This expression for the disc amplitude agrees after a rescaling of 
the cosmological constant with the amplitude 
$W^{(eu)}_\L (X)$ from 2d Euclidean 
quantum gravity, 
\beq\label{top21}
W_\L (X) = (X -\oh \SL ) \sqrt{X+\SL}.
\eeq
 With $W_\L(X)$ substituted into
\rf{top16}, the resulting equation is familiar from the usual
theory of 2d Euclidean quantum gravity, where it has been
derived in various ways \cite{kkmw,ik,watabiki},
with $T$ playing the role of {\em geodesic distance} between 
the initial and  final loop. 

Before showing that the anomalous scaling of the proper time --
once baby universes are allowed -- 
leads to an intrinsic fractal space-time dimension
of four (rather than two), let us comment 
on the difference between the equations for
the amplitudes \rf{an5c}-\rf{an6} for 
$(\eta,\ep) =(-1,1)$ and $(\eta,\ep)=(3/2,1/2)$ respectively. 
In the first case there are no baby universes and 
eq.\ \rf{an5c} entails that only {\em macroscopic loops} at a 
distance $T$ from the entrance loop are important (as illustrated 
in Fig.\ \ref{identity}). On the other hand, the term $\prt W_\L(Z)/\prt Z$
which describes the presence of these macroscopic loops is absent in
eq.\ \rf{an5}.
This is consistent with eq. \rf{an5a}, which shows explicitly 
that the length of the upper loop in Fig.\ \ref{identity} remains at
the cut-off scale, and therefore can never become macroscopic.
It is also consistent with the abundance of baby universes, since  
at any point in space-time the probability 
for creating a little ``tip'' of cut-off size 
will dominate. At the same time, the right-hand side of eq.\ \rf{an5c}, 
that is, eq.\ \rf{an7}, will play no role when $1 < \eta <2$, 
being simply equal to a constant. This latter property is 
satisfied automatically, as can be seen by using an equation 
analogous to \rf{top16} for the exit instead of the entrance loop. 
Thus eq.\ \rf{an7} becomes proportional to  
\beq\label{last}
\int_0^\infty \d T \frac{\prt}{\prt T} \;G(X,L_2 \equ 0;T) = 
\mbox{const.},
\eeq
proving our previous assertion. 
 
\subsection{The fractal dimension of Euclidean 2d 
gravity}\label{fract}

If we allow for baby universe creation,
the fractal structure of space-time 
is determined by \rf{top16} and \rf{top21}, where $T$ is 
the time separating the entrance and exit loops. 
As already mentioned, these are exactly the equations governing
Euclidean quantum gravity, if one replaces $T$ by
the geodesic distance between the two loops. 

One can solve eq.\ \rf{top16} in the same way as its Lorentzian 
analogue \rf{32} was solved by \rf{33}-\rf{35}.
We have
\beq\label{eu1}
G_\L (X,Y;T) = \frac{W_\L (\bar{X}(T;X))}{W_\L(X)} \;
\frac{1}{\bar{X}(T;X) +Y},
\eeq
where $\bar{X}(T;X)$ is the solution to the characteristic equation
\beq\label{eu2}
\frac{d\bar{X}}{dT} = -W_\L(\bar{X}),~~~~~\bar{X}(T=0)=X.
\eeq
This equation can be solved in terms of elementary functions.
In particular, one finds
\beq\label{eu2a}
\bar{X}(T;X=\infty) \propto \SL \, \coth^2 \Big( \a
\L^{1/4} T\Big),~~~~\a=\oh \sqrt{3/2}.
\eeq 
We may now define a {\it two-point function}
by contracting the entrance loop to a point and 
closing the exit loop by the disc amplitude. 
This is shown in Fig.\ \ref{identity},
except that now the entrance loop must be contracted\footnote{
One cannot {\it a priori} contract 
both loops, since the two points in the two-point function are
separated by a geodesic distance $T$, and there may be many points 
at distance $T$ from the entrance point, as shown in Fig.\ \ref{identity}.
They will in general form several connected loops. 
However, after solving the model, it turns out that one {\em can}  
just contract the exit 
loop and obtain the two-point function. The reason is that $w(x,g)$
contains a non-universal part, which again implies that a typical 
disc amplitude in Fig.\ \ref{identity} will be of cut-off size.
This is precisely the contents of eq.\ \rf{an5}, whose
non-scaling part is given by $\prt w_{ns}/{\prt z}|_{z=x_c}$.}.
The two-point function $G_\L(T)$ in Euclidean gravity 
can be interpreted as the (unnormalized) average 
over all geometries with two marked points which are separated 
by a geodesic distance $T$. 
At the discretized level it is defined by an equation 
analogous to \rf{an1}, but without summing over $t$,
\beq\label{eu3}
G(t;g) = \sum_l G(l_1,l;g;t)\; l\; w(l,g).
\eeq
As in eq.\ \rf{an4}, the continuum limit will be dominated by 
the non-scaling part of $w(l;g)$, i.e.\ by small $l$,  
and the continuum two-point function is simply given by
\beq\label{eu4}
G_\L (T) \sim G_\L (L_1\equ 0,L_2\equ 0;T).
\eeq
Using \rf{eu1} and the Laplace transforms of $G_\L$ and $W_\L$ we obtain
\beq\label{eu5}
G_\L(T) = \oint \frac{\d X}{2\pi i} \oint \frac{\d Y}{2\pi i}
\; \frac{W_\L (\bar{X}(T;X))}{W_\L(X)} \; 
\frac{1}{\bar{X}(T;X)+Y}.
\eeq
With the help of the characteristic equation \rf{eu2} this leads to
\beq\label{eu6}
G_\L (T)  \sim \frac{\d}{\d T} \oint \frac{\d X}{2\pi i} 
\;\frac{\bar{X}(T;X)}{W_\L(X)}.
\eeq 
The contour of integration can be deformed to infinity 
and we obtain from \rf{eu2a} that
\beq\label{eu7}
G_\L(T) \sim \L^{3/4} \;\frac{\cosh{\a \L^{1/4} T}}{\sinh^3 \a \L^{1/4} T}.
\eeq     
This two-point function may be viewed as the partition function 
for universes with two marked points separated by a geodesic distance
$T$. Since we wanted to solve {\it Euclidean} gravity, this is
our final result. Given \rf{eu7}, we
can calculate the average space-time volume of 
such a universe,
\beq\label{eu8}
\la V \ra_T = T^4 F\Big(\L^{1/4} T\Big),~~~~F(0) > 0.
\eeq  
The function $F$ is again expressible in terms 
of elementary functions and one finds
\beq\label{eu9}
\la V\ra_T \approx T^4~~~{\rm for}~~~ T < 1/\L^{1/4}.
\eeq
This formula shows that the intrinsic fractal dimension of 
2d Euclidean quantum space-time is four, whereas a similar derivation 
in the case of Lorentzian gravity yields \cite{al}
\beq\label{eu10}
\la V\ra_T \approx T^2 ~~~{\rm for}~~~~T< 1/\L^{1/2}.
\eeq
The discrepancy in dimension can be explained purely in
terms of the baby universe structure:
{\it since at each point of the two-dimensional Lorentzian surface 
a baby universe can branch off, the resulting fractal 
Euclidean space-time has twice the intrinsic dimension}.

\section{Euclidean quantum gravity}\label{euclidean}     

\subsection{Some generalities}

In the last section we arrived at the 2d Euclidean 
gravity theory through an ``extension'' of the Lorentzian model.
Euclidean quantum gravity can of course be defined independently,
as the quantization of classical gravity on the space of all
Riemannian metrics (of positive definite signature)
instead of the space of (indefinite-signature) Lorentzian metrics.
In two dimensions, Euclidean gravity has a well-defined 
continuum path-integral formulation. Choosing a conformal
gauge-fixing leads to the so-called Liouville 
gravity. Certain aspects of this theory can be solved 
by a bootstrap approach. In higher dimensions the path-integral 
approach to Euclidean quantum gravity is problematic, since
the Einstein-Hilbert action is unbounded from below. 
There are various ways of tackling this problem:
analytically continuing the unstable modes, adding stabilizing
higher-derivative terms to the action, or defining the theory 
non-perturbatively via a lattice regularization, 
such that the action is bounded for any finite lattice volume. 

One example of the latter is the dynamical-triangulations
approach, which has the added bonus of being exactly soluble
by combinatorial methods in two dimensions. Its continuum limit  
agrees with continuum Liouville quantum gravity, wherever the
two formulations can be compared.
We are in fact in the unusual situation 
that the lattice approach can address and answer more 
questions than the continuum methods. We will not review the 
combinatorial
approach here since the results (for pure gravity) were 
already obtained in Section \ref{topology}, starting from Lorentzian 
gravity. For
a detailed description we refer to \cite{book}, chapter 4.  
The generalization of this Euclidean lattice path integral to 
higher dimensions is straightforward
and shares two virtues with the 2d case: calculating the 
partition function for gravity
is again turned into a combinatorial problem, 
and the model is well-suited for numerical simulations.
We have by now a good qualitative understanding of the
phase structure of Euclidean dynamically-triangulated gravity 
in $d=3,4$, although complete analytical 
solutions of the discretized models are still missing. 
However, the combinatorial nature of the partition
function gives us some hope 
that progress can be made also in these cases.

\subsection{Dynamical triangulations}\label{dyna}

This lattice approach shares many elements with
lattice regularizations of ordinary quantum field theory. 
The main difference lies in the fact that the
geometric degrees of freedom become dynamical and 
the lattices are therefore no longer part of the inert background
structure.
The geometric quantum fluctuations must be taken properly into 
account when building discretized models of matter 
interacting with quantum gravity.
The field-theoretical, non-perturbative Euclidean path integral 
of such a theory takes the general form
\beq\label{part}
Z[G,\L,\{\beta_i\}] = \int\cD [g_{\mu\nu }]\cD \phi_i\ 
e^{-S_{matter}[g_{\mu\nu },\phi_i;\{\beta_i\}]-S_{EH}[g_{\mu\nu };G,\L]},
\eeq
where the integration is over equivalence classes of metrics 
$[g_{\mu\nu}]$ and matter fields $\phi_i$.
The action contains a matter part $S_{matter}$, depending on a set
of matter couplings $\{\beta_i\}$, and a purely geometric part,
given by the Einstein-Hilbert action with a cosmological term,
\beq\label{act}
S_{EH}[g_{\mu\nu };G,\L]=\frac{1}{16\pi G}\int d^dx\
\sqrt{\det g}\, (2\L-R).
\eeq
In \rf{part} we have omitted possible boundary terms. 
Except in $d=2$, expressions of the kind \rf{part} have remained formal, 
due to the absence of a suitable diffeomorphism-invariant
integration measure. 
The lattice formulation is an attempt to remedy this situation,
by using an intermediate regularization of the non-perturbative
path integral \rf{part} (see \cite{ajw1,book} for reviews).
Defining a discrete regularization consists of several steps:
\begin{itemize}
\item[$\bullet$] a discretization of the individual metric manifolds, 
together
with a definition of discretized geometric ``observables'',
such as lengths, volumes and (scalar) curvature. These are
necessary for obtaining a discretized version of the action,
and for analyzing the physical properties of the theory in terms
of scaling relations and correlation functions.
\item[$\bullet$] a suitable choice of an integration measure on the space of
discretized geometries (i.e. equivalence classes of metrics), 
such that the discrete path integral converges.
\item[$\bullet$] a discretization of the matter sector, which will be closely
related to standard lattice-regularizations in field theory. 
\end{itemize}

Let us now describe the dynamical-triangulations regularization 
of Euclidean quantum gravity. 
It consists in replacing the $d$-dimensional Riemannian metric
continuum manifold by a simplicial manifold constructed from
equilateral $d$-dimensional simplices of (geodesic) edge length $a$. 
(A simplex is a point in $d=0$, an edge in $d=1$, a triangle in
$d=2$, a tetrahedron in $d=3$, etc.)
Using Regge's prescription \cite{regge}, all quantities
can be expressed as functions of the squared edge lengths.
For example, the curvature depends on local deficit angles,
which in turn are expressible in terms of edge lengths.
A simplicial complex is obtained
by gluing together $d$-simplices pairwise 
along $(d-1)$-dimensional faces (which are themselves
$(d-1)$-simplices). 
Since, unlike in Regge calculus, our edge lengths are not variable,
all $d$-simplices have the same size, and
the total volume of the simplicial complex is simply proportional 
to the number $N_d$ of such cells.
Each $d$-simplex is built from simplices of
lower dimensionality. It contains $d+1$ 0-simplices (vertices), $(d+1)d/2$
1-simplices (links) etc. A lower-dimensional subsimplex is in general
shared by a number of $d$-simplices, called the {\it order} of the subsimplex.
A simplicial complex is a simplicial manifold if 
the neighbourhood of any $p$-simplex ($p < d$) has the topology of a 
$(d-p-1)$-dimensional sphere. 
Physically the manifold requirement may be viewed
as a regularity condition at the cut-off scale, which will be 
convenient to use in our construction.
The numbers $N_k$ of (sub-)simplices of dimension $k \le d$ are not
independent, but (due to the regularity requirement) must
satisfy a set of so-called Dehn-Sommerville relations, namely,
\beq\label{DS}
N_i=\sum_{k=1}^d (-1)^{k+d} \left( \begin{array}{c} k+1\\ i+1 \end{array}
\right) N_k,
\eeq
together with the Euler constraint
\beq\label{Euler}
\sum_{k=0}^d (-1)^k N_k = \chi.
\eeq
For fixed Euler number $\chi$ and $d=2$, all $N_k$ can be expressed 
as functions of the single variable $N_2$, say. For $d=3,4$, two of
the $N_k$ are independent.

A triangulation $T$ together with an assignment of geodesic edge
lengths and flat simplex interiors may be viewed as a piecewise linear 
manifold, and provides an explicit 
coordinate-independent representation of a metric manifold.
In this way each triangulation corresponds to a unique equivalence 
class of metrics (albeit of piecewise-linear, and not of differentiable
type).

When it comes to numerical simulations, it is often convenient to 
assign a label to each vertex. From the list of vertex labels
for all $d$-simplices the whole manifold can be reconstructed. 
Since the labels themselves have no physical meaning, the
labelling introduces a redundancy. Invariance under permutations
of the labels may loosely be regarded as a discrete analogue
of the diffeomorphism invariance of a differentiable manifold. 
Also the lower-dimensional subsimplices are characterized by their
vertex labels. Moreover, the regularity requirement implies that
we cannot have two
different (sub-)simplices with the same set of vertex labels.
For each triangulation $T$ with $N_0$ vertices
the number of different labellings equals $N_0 !/ C(T)$,
where $C(T)$ is the order of the automorphism group of $T$. 
We can therefore distinguish between
labelled triangulations $\tilde{T}$ and abstract unlabelled 
triangulations $T$.
As mentioned above, different $T$'s (with fixed topology)
correspond to different equivalence classes of piecewise linear metrics
assigned to the underlying manifold, and allow us 
to work directly with a reparameterization-invariant set of geometries.

Since the simplices are flat on the inside, curvature is located
(distributionally) at simplices of lower dimension. 
Circulating around a $(d-2)$-dimensional simplex, the contributing
angles will in general not add up to $2\pi$. The resulting deficit angle
depends on the number of simplices meeting at the subsimplex. 
We conclude that the curvature is concentrated at the $(d-2)$-dimensional 
subsimplices of the triangulation.
The Einstein-Hilbert action for a dynamically
triangulated manifold in dimension $d$ assumes the simple form
\beq\label{HED}
S_{EH} = \kappa_d N_d - \kappa_{d-2} N_{d-2},
\eeq
with the two dimensionless coupling constants $ \kappa_d$ and
$\kappa_{d-2}$.
As usual in lattice regularizations, the
lattice spacing $a$ has disappeared from the formulation and will
have to be reintroduced in the scaling limit. 
We are using the Einstein-Hilbert action because of its
simplicity; one could in principle consider also the inclusion of
higher-order curvature terms.

With each simplicial lattice described above 
one can associate a dual lattice,
whose vertices are located at the centres of the simplices of the 
original lattice. 
In a similar way we can associate to each $p$-simplex 
a dual object of dimension $d-p$. For example, the
dual links connect the centres of neighbouring simplices, and the dual of a 
$(d-2)$-simplex is a closed loop (a two-dimensional object) whose length
is equal to the number of $d$-simplices of the original lattice which share
the $(d-2)$-simplex. Since in $d$ dimensions
each simplex has $d+1$ neighbours, the dual lattices have the form of
graphs of a scalar $\bf{\phi}^{d+1}$-theory (that is, all their
vertices are $(d+1)$-valent), but
with a local $d$-dimensional topological structure. 

The simplicial structures described above possess a natural
notion of length for any path connecting two points, since
the equivalence class of metrics is uniquely determined. 
To simplify matters, we will only consider certain sets of
discretized paths on the simplicial manifold.
The first set is given by paths which run along the
links of the original simplicial lattice,
and the other by paths running
along the links of the dual lattice. In either case we may
define a distance between points on the lattice or its dual
as the number of edges of the shortest path connecting the two.
At first glance these definitions seem different from the
standard notion of a geodesic distance, but 
they all coincide in the scaling limit, up to trivial
numerical factors\footnote{The situation is the same as for a 
regular 2d quadratic lattice in flat space: if we are only allowed 
to connect vertices along the lattice links, the lattice 
distance between 
different lattice points can differ by as much as a factor $\sqrt{2}$ from 
the Euclidean distance in flat space. However, in the scaling limit 
the rotational symmetry of the original field theory will be
restored on the lattice and the two notions of distance will only
differ by an overall factor.}. 
Numerical tests of this assumption will be discussed below.

\subsection{The functional integral}\label{funct}

The association of triangulations with equivalence classes of metrics
motivates the use of the discrete sum over
$d$-dimensional triangulations as 
a discretized analogue of the diffeomorphism-invariant
integration measure in \rf{part},
\beq\label{measure}
\int \cD [g_{\mu\nu}] \to \sum_T {1 \over C(T)},
\eeq
where the sum is taken over unlabelled simplicial manifolds.
The need for including the symmetry factor $C(T)$
becomes apparent when one rewrites the right-hand side of \rf{measure} as
a sum over labelled triangulations $\tilde{T}$,
\beq\label{measure1}
\sum_T {1\over C(T)} \to \sum_{\tilde{T}} {1\over N_0(\tilde{T})!}.
\eeq
In order that a discretized path integral with this choice of measure
lead to a theory with a well-defined thermodynamic limit,
the number of triangulations with fixed volume $N_d$ should grow at most
exponentially with $N_d$ as $N_d \to \infty$. 
This is not the case unless we fix the space-time topology 
(usually to that of a sphere S$^d$); otherwise the growth is factorial. 
This property has been proven for $d=2$ 
and arbitrary topology \cite{benderetal} and for simply connected 
manifolds in $d=3,4$ \cite{carfora,acm}.

It is worthwhile pointing out that the simple choices 
\rf{HED} for the action and \rf{measure} for the measure
lead to a partition function of the form
\beq\label{countq}
Z(g_d,g_{d-2}) = \sum_{T} \frac{1}{C(T)} \; g^{N_d(T)}_d \, 
g_{d-2}^{N_{d-2}(T)},
\eeq  
where $g_{d}= -\log \k_d$ and $g_{d-2}= \log \k_{d-2}$.
Eq.\ \rf{countq} shows that the partition function is the generating 
function for the number of triangulations (of fixed topology) with 
given numbers $N_d$ and $N_{d-2}$ of $d$- and 
$(d-2)$-simplices. We thus reach the surprising conclusion that 
{\it even in dimension} $d>2$, quantum gravity can be formulated as a 
(relatively simple) counting problem.

\subsection{Inclusion of matter fields}\label{matter}

The discretization of matter fields coupled to 
dynamical triangulations is achieved by standard 
lattice field-theoretical methods.
The simplest types of matter fields one can study 
are either scalar fields or (Potts) spin fields, 
carrying a discrete space-time label.
One may also combine several fields of this type.
The matter fields can be located at the vertices of the triangulation or at 
the centres of the $d$-simplices (that is, at the dual vertices). 
The interactions are typically of the form of
nearest-neighbour interactions, where 
the ``nearest neighbours'' are the vertices that are one link length
(or one dual link length) away from the original vertex.
We expect these two formulations to become equivalent in the scaling limit.
Some typical examples of matter actions are
\beq\label{ising}
S_I = {\beta\over 2} \sum_{\{ij\}} (\delta_{\sigma_i \sigma_j}-1), 
\eeq 
where the $\sigma_i$ are a set of Potts spins, or the Gaussian action
\beq\label{gaus}
S_g = \sum_{\{ij\}} (\phi_i - \phi_j)^2
\eeq
for a massless scalar field $\phi$.
Note that we did not need to include a coupling constant
in front of the action \rf{gaus}. 
The massless scalar field can always be
rescaled by a factor, which can then be absorbed by 
a redefinition of the coupling constants of the geometric sector.
The coupling of Ising 
spins (Potts spins $\sigma_i$ with $i=1,2$) and Gaussian fields
to the 2d Lorentzian gravity model proceeds in a manner
identical to \rf{ising} and \rf{gaus}.

In higher dimensions we will also consider the coupling to
gauge fields. 
As usual these are associated with the one-dimensional edges
of the triangulation, and there are again two alternative formulations,
depending on whether the links or the dual links are used.
The gauge field action is more complicated and we postpone its
discussion to Section \ref{higherd}.
The inclusion of fermionic degrees of freedom
on a random manifold remains an open problem, particularly in higher
dimensions. It requires the definition of a spin connection on a simplicial
manifold (c.f. \cite{reggestuff}). The problem was solved 
in $d=2$ \cite{andre-jurek-spin}. In this case  
one can prove that a system of Wilson fermions 
on a triangulated manifold can be ``bosonized'' and represented as
a system of Ising spins on the manifold \cite{migdal,andre-jurek-spin}.    

The discretized path-integral measure contains also a sum
over matter fields. 
In the case of spin variables, the sum is simply taken over
all possible spin configurations.
For continuous fields like the scalar fields above,
one may consider non-trivial integration 
measures which introduce an additional 
coupling to geometry. This possibility does not exist
in field theories on fixed, regular lattices, where such a dependence
is always trivial. It leads to some subtleties in the case of
gauge fields, as we will discuss later.

The simplest and most extensively studied example 
of a dynamically triangulated theory is that of Euclidean gravity
on a two-sphere.
The fundamental building blocks in this case are equilateral
triangles (2-simplices). Triangles are glued together pairwise along
edges (1-simplices), and each triangle has exactly
three neighbours. The dual lattice is thus equivalent to a planar
$\bf{\phi}^3$-diagram. The curvature is localized at the 
vertices (0-simplices) which in general are shared by 
many triangles, each contributing $\pi/3$ to the total
angle around the vertex. The regularity requirement
introduced above prohibits configurations where a vertex is 
its own neighbour or where two vertices are connected by more than
one link (in other words, closed loops of link length one or two
are forbidden).

The Lorentzian gravity model introduced in Section \ref{model}
may be viewed as a restricted version of the dynamically
triangulated Euclidean model, since the triangulations contributing
to the Lorentzian state sum constitute a subset of those
appearing in the Euclidean system defined above.
Recall also our construction of Euclidean from Lorentzian
gravity in Section \ref{baby}, by allowing for additional
baby-universe branching. Again the set of all such geometries
is a subset of all 2d simplicial manifolds, but both continuum
theories coincide. What is at work in this latter case is 
``universality'', which ensures that
the continuum limit is to a large extent independent of the 
short-distance details characterizing the class of
triangulations we sum over.
The universality properties of Euclidean 2d quantum gravity 
are well studied. For example, one may relax the manifold
regularity condition to obtain a much larger class of simplicial
complexes, whose continuum limit is still Euclidean quantum
gravity.
Only drastic modifications, like the suppression of baby universes,
can move the system to a different universality
class with a different critical behaviour and therefore a
different continuum limit.
Also around the fixed point leading to Lorentzian 2d gravity
one finds an independence of short-distance details. 
Universality with respect to a change of fundamental building
blocks and the inclusion of higher curvature has been demonstrated 
in \cite{lotti}. It is also encountered in a recently developped
procedure for obtaining Lorentzian from Euclidean quantum
gravity by removing baby universes \cite{ackl}. There one ends
up with a generalized class of triangulations (compared to
the original Lorentzian model), but the continuum limit is
still the same.

\section{Numerical setup}\label{numerical}

\subsection{Monte Carlo method and ergodic moves}\label{monte}

Even in two dimensions, there are a number of issues
of the matter-coupled theory that presently can only be addressed
by numerical methods. 
Lorentz\-ian gravity coupled to Ising spins has not yet
been solved analytically. 
In matter-coupled Euclidean 2d quantum gravity,
analytical considerations have not yet led to a determination of
the fractal dimension of space-time (there are various suggestions
leading to different answers), nor do we know
what happens beyond the infamous $c=1$ barrier, where analytical 
calculations break down. 
In higher dimensions, we do not even have analytical 
solutions of the pure-gravity models. 
In all of these situations, numerical simulations of the 
systems come in handy. They can answer specific 
questions and lead to unexpected results which in turn 
can inspire further analytical work.  

Numerical simulations of simplicial gravity have been the subject
of a number of reviews (see \cite{andre,numreview} for annual updates 
and \cite{numreview1} for more information on the computer codes
used in the programs).  
Here we will only sketch the methods and use the simplest
case of 2d Euclidean gravity as an illustration. 
Most of what we will have to say carries over 
to 2d Lorentzian gravity with only minor modifications.

As explained in the previous section, the discretized theory can be
described by the partition function
\beq\label{partd}
Z = \sum_{\tilde{T}} \frac{1}{N_0(\tilde{T})!}
e^{-S_{EH}(\tilde{T})}\sum_{\phi_i}
e^{-S_{matter}(\phi_i)},
\eeq
where  $S_{EH}$ is the discretized Einstein-Hilbert action
\rf{HED},
and the first sum is taken over all labelled triangulations of
fixed spherical topology. 
For $d=2$, the geometric part of the action simplifies and takes the form
\beq\label{HE2d}
S_{EH}=\m N_2,
\eeq 
up to an additive constant proportional to
the Euler number $\chi$, with $\m= (\kappa_2-\kappa_0)/2$. 
In numerical simulations, it is simpler to use the labelled instead
of the unlabelled triangulations.
The partition function \rf{partd} is the analogue of the grand canonical
partition function in statistical mechanics. The cosmological constant
$\kappa_d$ plays the role of a thermodynamic potential for the number
of simplices. For general $d$, we may rewrite \rf{partd} as
\beq\label{partd1}
Z = \sum_{N_d} e^{-\kappa_d N_d} Z_{N_d},
\eeq
where $Z_{N_d}$ can be interpreted as the partition function at
fixed volume. 
A gravity-matter configuration $C$ is uniquely specified 
by a geometry in the form of a labelled triangulation and by the values
of all matter field variables. Each configuration 
enters the statistical sum with a probability proportional 
to $e^{-S(C)}$, 
where $S(C)=S_{EH}+S_{matter}$. As usual in statistical 
mechanics, physical information
can be obtained by measuring the averages of various 
operators in this ensemble. The fact that each configuration
has a real positive weight makes it possible to study the system
\rf{partd} by Monte Carlo methods. The goal is to construct
a numerical ``generator'' which
produces configurations $C$ with a probability $P(C) \sim e^{-S(C)}$.

Except for very few cases, where a direct generation of all configurations
is possible (e.g. for a conformal charge $c=-2$ in 2d),
the standard way of obtaining
such a generator is by means of a stochastic process (a Markov chain),
which can be regarded as a random walk in configuration space. 
Since the random walk
takes place on a computer, each step corresponds to the real time needed
to perform such an operation. The stochastic process is
defined by a function $W(C \to C') \ge 0$, giving the probability
for a transition from a configuration $C$ to $C'$ in one step. 
Assuming for simplicity 
that the configuration space is discrete, we have a normalization condition
\beq\label{norm}
\sum_{C'} W(C \to C')=1.
\eeq
In most cases $W(C \to C')$ is chosen to vanish outside some
``neighbourhood'' of $C$. Starting from an initial state $p_0(C)$,
the system after $n$ steps is characterized by a probability
distribution $p_n(C)$, where 
\beq\label{next}
p_{n+1}(C) = \sum_{C'} p_n(C') W(C' \to C).
\eeq

The transition probability $W$ must satisfy two basic requirements,
\begin{itemize}
\item[1.] {\it ergodicity:} any two configurations can be joined
by a finite number of steps, and
\item[2.] {\it detailed balance:} the condition 
\beq\label{db}
P(C) W(C \to C') = P(C') W(C' \to C),
\eeq
relating $W(C \to C')$ to $W(C' \to C)$.
\end{itemize} 
It follows from \rf{db} that $W(C \to C')$ and  $W(C' \to C)$ 
are either both zero or both non-zero in which case they satisfy
\beq\label{dbcimp}
\frac{W(C \to C')}{W(C' \to C)} = \frac{P(C')}{P(C)}.
\eeq
These two requirements guarantee that
the stochastic process has a unique asymptotic probability distribution
$p_{\infty}(C) \sim P(C)$, which is the only eigenstate of the transition
matrix $W$ with eigenvalue 1,
\beq\label{asympt}
\sum_{C'} p_{\infty}(C') W(C' \to C) = p_{\infty}(C).
\eeq
All other eigenvalues are strictly smaller than 1. 
This implies that -- independent of
the initial configuration -- the system will reach the 
asymptotic distribution after infinitely many steps. 
Note that the asymptotic distribution has
the desired probability distribution. The rate at which the system
approaches this limiting distribution depends on 
the other eigenvalues of $W$.
The contributions from other eigenstates decay exponentially with the number
of steps.
The second-largest eigenvalue $\l_1$ provides us with an
estimate of the autocorrelation time $T=-1/\log \l_1$.
When the number $n$ of steps is $\gg T$,
we can assume that the distribution is asymptotic.
Typically the autocorrelation time behaves like $T \sim f^{\delta}$, 
where $f$ counts the number of degrees
of freedom of the system and $\delta$ is a dynamical exponent, 
depending on the details of the algorithm.

In a practical implementation the system starts from some configuration $C_0$.
During the first step, it changes to $C_1$ with probability 
$W(C_0 \to C_1)$, or remains at $C_0$ if the change is not accepted. 
After $n \gg T$ steps it reaches
a configuration $C_n=C^{(1)}$ with a probability proportional to $P(C^{(1)})$. 
This configuration is the starting point of a new process, 
during which another sufficiently
large number of steps is performed and a new configuration $C^{(2)}$ 
generated. Repeating this process we create a (finite) set of 
configurations $\{ C^{(1)},\dots,C^{(N)}\}$, where $N$ depends on
the computer time we spend on the project. The average of any operator
in the ensemble of configurations defined by the probability distribution 
$P(C)$ is approximated by an average over this finite sample of 
``typical'' configurations.

The requirements listed above by no means define the stochastic
process uniquely. 
We are interested in efficient algorithms which minimize $\delta$,
and can produce a large number of independent configurations in
the shortest possible time. There is no simple way to guess 
at the outset whether an algorithm is efficient or not. Each problem must be
treated individually and the autocorrelation time measured numerically.
There are some general guidelines which one usually
follows when creating a new algorithm. As discussed above, the efficiency
depends on the choice of the transition probabilities $W(C \to C')$.
We would like the algorithm to have a high ``mobility'', that is,
a high probability that a configuration of the system will change 
at each step. This means that
the set $\{ C' \}$ of configurations which can be reached from a given $C$
should be limited. If it is too large, each transition probability
will be small, implying that the configuration will most likely
not change. On the other hand, the set \{$C'$\} must be large enough
to ensure ergodicity.

The detailed-balance condition \rf{dbcimp} implies that if both 
transitions are to be reasonably probable, we must choose the
set \{$C'$\} such that its elements have similar probabilities to
that of $C$ (equivalently, have a small action difference
$|S(C)-S(C')|$).  
On a fixed lattice, small differences in the action are usually 
realized by considering at each step only local changes in the 
field variables, for instance, changing
only one variable, while keeping all the others fixed. 
When a more complicated
change is attempted, the action difference is in general 
large and proportional to the volume of the system.
Local changes are not very efficient when the typical fluctuations
are long-ranged, as happens close to a continuous
phase transition. Creating a Monte Carlo algorithm which is at the same
time ergodic and has a reasonably small autocorrelation time even in
the critical region is an art.

The first step in constructing an algorithm for simplicial gravity is
to define a method of coding the configurations. From
a numerical point of view it is natural to work with labelled rather
than unlabelled triangulations, because otherwise it is almost
impossible to keep track of the (dynamical) connectivity. 
To code the geometric structure of such a configuration it
is in principle sufficient to have a list of the vertex labels 
of all simplices (of all dimensions) of the triangulation.
Two vertices are neighbours if they belong to the same simplex. From
this list we can reconstruct the complete geometry of the
system. The fact that we have a list of simplices means in
practice that we have also assigned labels to the simplices. Two simplices
are neighbours if they share a $(d-1)$-simplex or, equivalently,
$d$ vertices. 
In addition, we can make a list of all subsimplices and count their
order. The process of reconstruction may still be complicated and 
it is often useful to keep at each step even more information, in
the extreme case the lists of all subsimplices. The
more information we keep, the easier it is to reconstruct local
properties of the geometry. However, it also means that during
geometry updates more data will have to be changed.
  
The next step consists in constructing the stochastic
process described above. In our case this amounts to 
defining a set of ``moves'', which
connect configurations with different geometry and matter content. 
Usually these two changes are performed separately: 
changes of the matter fields
are performed for a fixed geometric configuration, by
techniques very similar to those used for fixed lattices. 
For spin systems
there exist highly effective cluster algorithms \cite{cluster}
with very short autocorrelation times. Cluster algorithms are
special in that they permit global changes of the spin configurations.
For Gaussian fields, no such algorithms exist and the field variables 
are usually updated sequentially. (There
are some methods to shorten their autocorrelation time.) 
Generally speaking, the
autocorrelation times for Gaussian fields are much longer than
those for spin systems. It is therefore necessary to
repeat the updating of all fields on the manifold several times 
in order to obtain independent field configurations. 

In addition, we must define a set
of geometric moves which generate local changes of the geometry,
while preserving its underlying topology. Such moves are necessary even
in the absence of matter.
``Local changes'' means that only a small region of the manifold is 
affected at each step. Again, it is important that the set of moves 
be ergodic in the space of all possible triangulations. 
For simplicial manifolds, one possible choice is given by
the so-called Alexander moves \cite{alexander}. In practical
applications most algorithms for a $d$-dimensional geometry are based on 
a finite subset of moves containing $d+1$ operations. We will describe
this set of operations in the case of 2d and postpone
its generalization to higher $d$ to Section \ref{higherd}. 

The first operation is called a flip and involves two
neighbouring triangles. The triangles are denoted by their vertex 
labels, $\{123\}$ and $\{124\}$, and share the link $\{12\}$.
It is important that the four labels are all distinct 
(excluding $\{3\} =\{4\}$, say) and that the vertices $\{3\}$ and $\{4\}$
are not connected by a link. The flip consists in replacing
this configuration by the two triangles $\{341\}$ and $\{342\}$. 
In other words, the link $\{12\}$ is ``flipped'' to the link $\{34\}$. 
The restrictions imposed above 
guarantee that this move does not produce a pathological triangulation.
The inverse of the flip operation is again a flip.
Note that the flip move takes place entirely inside the link loop
$\{1324\}$, whose boundary it leaves unchanged. 
Since two simplices are replaced by two others, the flip is
sometimes called a $(2,2)$-move. 
A flip can be performed almost everywhere in the manifold, 
giving rise to a different labelled manifold whose
connectivity is changed locally. If matter fields are present at
the vertices, one usually assumes that their values are unchanged.
The transition weights $W(C \to C')$ and $W(C' \to C)$ for a flip and 
its inverse can easily be computed from the detailed-balance condition. 
We will not be more specific here, since
this depends on the details of the geometric coding 
(in particular, on how links are selected).
A flip move leaves the numbers of triangles and vertices unchanged.
  
\begin{figure}
\unitlength=1mm
\linethickness{0.4pt}
\begin{picture}(70.00,50.00)(-20,5)
\bezier{80}(47.00,40.00)(59.00,40.00)(67.00,40.00)
\bezier{72}(47.00,40.00)(53.00,49.00)(57.00,55.00)
\bezier{72}(57.00,55.00)(62.00,48.00)(67.00,40.00)
\bezier{80}(10.00,40.00)(22.00,40.00)(30.00,40.00)
\bezier{72}(10.00,40.00)(16.00,49.00)(20.00,55.00)
\bezier{72}(20.00,55.00)(25.00,48.00)(30.00,40.00)
\bezier{80}(10.00,15.00)(22.00,15.00)(30.00,15.00)
\bezier{72}(10.00,15.00)(16.00,24.00)(20.00,30.00)
\bezier{72}(20.00,30.00)(25.00,23.00)(30.00,15.00)
\bezier{68}(10.00,15.00)(15.00,8.00)(20.00,1.00)
\bezier{68}(30.00,15.00)(25.00,8.00)(20.00,1.00)
\bezier{72}(45.00,15.00)(54.00,21.00)(60.00,25.00)
\bezier{72}(45.00,15.00)(53.00,9.00)(60.00,5.00)
\bezier{80}(60.00,5.00)(60.00,15.00)(60.00,25.00)
\bezier{72}(60.00,25.00)(68.00,20.00)(75.00,15.00)
\bezier{72}(75.00,15.00)(68.00,10.00)(60.00,5.00)
\bezier{40}(20.00,45.00)(20.00,50.00)(20.00,55.00)
\bezier{44}(20.00,45.00)(16.00,43.00)(10.00,40.00)
\bezier{44}(20.00,45.00)(25.00,43.00)(30.00,40.00)
\put(33.00,47.00){\vector(1,0){10.00}}
\put(43.00,47.00){\vector(-1,0){10.00}}
\put(33.00,15.00){\vector(1,0){9.00}}
\put(42.00,15.00){\vector(-1,0){9.00}}
\end{picture}
\caption[flip]{A set of three moves which is ergodic
in the class of two-dimensional
triangulations of fixed topology. The first diagram shows
the $(3,1)$-move and its inverse.
The second diagram shows the $(2,2)$- or flip move.
By itself, this move is ergodic in the class of triangulations 
of fixed volume $N_2$ and topology.}
\label{flip}
\end{figure}
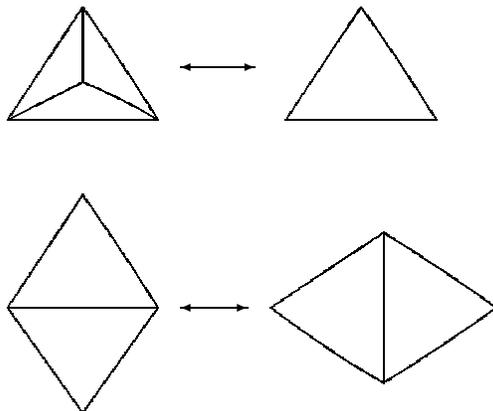

The second move adds a new vertex $\{4\}$ inside a triangle
$\{123\}$. This move replaces the old triangle by the three  
triangles $\{124\}$, $\{234\}$ and $\{314\}$, and leaves the
manifold outside the ``boundary'' of $\{123\}$ unaffected. 
It is also known as a $(1,3)$-move. 
The new vertex is of order three, since it is shared by
three triangles. The third move is the inverse
of the $(1,3)$-move. It is a $(3,1)$-move where a vertex of order three is
removed, and the three triangles which share it are replaced by a
single triangle. 
The $(1,3)$-move can be performed on each triangle of a manifold,
but its inverse needs a vertex of coordination number three.
It is obvious that in both cases the configurations before and
after the move are two different labelled triangulations. 
If matter fields are present, a new field must be created at the 
new vertex generated during a $(1,3)$-move. 
This must be taken into account when the transition weights are
calculated. 
The three types of moves are depicted in Fig.\ \ref{flip}.

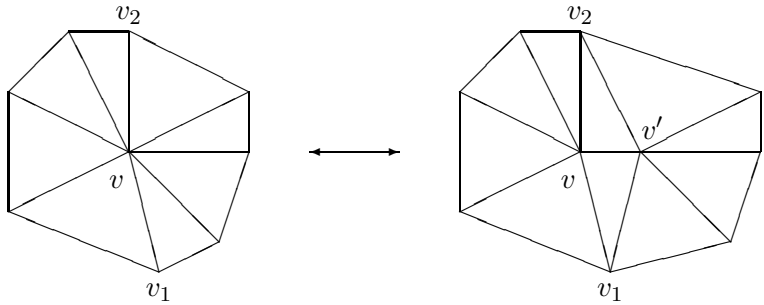
\begin{figure}
\unitlength=0.8mm
\linethickness{0.4pt}
\begin{picture}(135.00,62.00)(-10,10)
\put(30.00,40.00){\line(2,1){20.00}}
\put(50.00,50.00){\line(-2,1){20.00}}
\put(30.00,60.00){\line(0,-1){20.00}}
\put(30.00,40.00){\line(-1,2){10.00}}
\put(20.00,60.00){\line(1,0){10.00}}
\put(30.00,40.00){\line(-2,1){20.00}}
\put(10.00,50.00){\line(1,1){10.00}}
\put(30.00,40.00){\line(-2,-1){20.00}}
\put(10.00,30.00){\line(0,1){20.00}}
\put(30.00,40.00){\line(1,0){20.00}}
\put(50.00,40.00){\line(0,1){10.00}}
\put(30.00,40.00){\line(1,-1){15.00}}
\put(45.00,25.00){\line(1,3){5.00}}
\put(30.00,40.00){\line(1,-4){5.00}}
\put(35.00,20.00){\line(2,1){10.00}}
\put(35.00,20.00){\line(-5,2){25.00}}
\put(60.00,40.00){\vector(1,0){15.00}}
\put(75.00,40.00){\vector(-1,0){15.00}}
\put(115.00,40.00){\line(2,1){20.00}}
\put(105.00,60.00){\line(0,-1){20.00}}
\put(105.00,40.00){\line(-1,2){10.00}}
\put(95.00,60.00){\line(1,0){10.00}}
\put(105.00,40.00){\line(-2,1){20.00}}
\put(85.00,50.00){\line(1,1){10.00}}
\put(105.00,40.00){\line(-2,-1){20.00}}
\put(85.00,30.00){\line(0,1){20.00}}
\put(115.00,40.00){\line(1,0){20.00}}
\put(135.00,40.00){\line(0,1){10.00}}
\put(115.00,40.00){\line(1,-1){15.00}}
\put(130.00,25.00){\line(1,3){5.00}}
\put(105.00,40.00){\line(1,-4){5.00}}
\put(110.00,20.00){\line(-5,2){25.00}}
\put(105.00,60.00){\line(3,-1){30.00}}
\put(110.00,20.00){\line(4,1){20.00}}
\put(105.00,40.00){\line(1,0){10.00}}
\put(105.00,60.00){\line(1,-2){10.00}}
\put(115.00,40.00){\line(-1,-4){5.00}}
\put(28.00,35.00){\makebox(0,0)[cc]{$v$}}
\put(35.00,18.00){\makebox(0,0)[ct]{$v_1$}}
\put(30.00,62.00){\makebox(0,0)[cb]{$v_2$}}
\put(105.00,62.00){\makebox(0,0)[cb]{$v_2$}}
\put(110.00,18.00){\makebox(0,0)[ct]{$v_1$}}
\put(103.00,35.00){\makebox(0,0)[cc]{$v$}}
\put(117.00,44.00){\makebox(0,0)[cc]{$v'$}}
\end{picture}
\caption[split]{The point-splitting moves constitute an 
alternative set of ergodic moves.}
\label{split}
\end{figure}

The three operations just described leave the manifold topology 
unchanged. They form an ergodic set, which means that any two 
simplicial manifolds
of the same topology can be related by a finite sequence of moves. 
In 2d one can construct such a sequence explicitly. 
The set of ergodic moves is not unique, and alternative sets
of local moves have been used in applications.
For example, a point-splitting algorithm is described in 
\cite{book}. The point-splitting move and its inverse 
(illustrated in Fig.\ \ref{split}) change the volume $N_2$ by 
$\pm 2$. The $(1,3)$- and $(3,1)$-moves are
special cases and the $(2,2)$-move can be realized as a sequence 
of two point-splitting moves. 

\begin{figure}
\centerline{\hbox{\psfig{figure=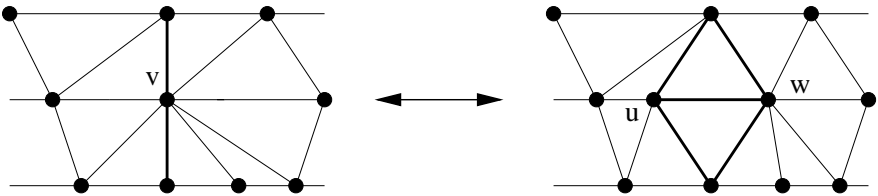,height=2.5cm,angle=-90}}}
\caption[move]{The move used in the Monte Carlo updating of 
a 2d Lorentzian geometry.}
\label{lorentzmove}
\end{figure}

Our discussion so far suggests that we may set up a numerical
simulation by generating configurations according to the
probability distribution \rf{partd}, 
and use the resulting sample of configurations
to measure the quantities of interest. 
However, this is not really feasible,
since the system described by \rf{partd} is an open system 
in the sense that arbitrarily large configurations
may be produced 
by using one of the sets of geodesic moves described above.
In practice, we must limit this size because of
the obvious memory restrictions of a computer. 
A simple solution is
to generate a set of configurations of fixed volume $N_2$, and repeat
the experiment for various values of $N_2$. 
This is what one typically does 
in numerical simulations of field theory. 
For the case of Euclidean 2d triangulations we are
particularly lucky, since the (2,2)-flip move is already by itself 
ergodic in the space of
all triangulations of fixed volume $N_2$, considerably
simplifying the computer simulations. 
Unfortunately there is no similar result in higher dimensions; 
we will describe later the method used in this case.
The flip move cannot be used in numerical simulations of the Lorentzian
model described in Section \ref{model}, since it is not compatible with
the causal structure. One uses instead a version of the
point-splitting move (Fig.\ \ref{lorentzmove}) to update the
geometry (see \cite{aal1} for details). 

We conclude this section by describing another type of move
which is associated with a large change in matter and geometry
(while keeping $\Delta S$ small). It is motivated by the baby-universe 
structure of Euclidean simplicial gravity. 
Let us consider a two-dimensional triangulated
manifold with the topology of a disc. 
It is characterized by its volume $n$ and boundary length 
$l$. For a random geometry these two numbers are not related 
since for any finite $l$, $n$ can become arbitrarily large.
If we have two such discs, of volume $n$ and 
$N-n$, but with identical boundary lengths, we can identify
their boundaries to create a closed simplicial complex 
of spherical topology and volume $N_2=N$. 
The smaller component is the baby universe and the larger one the
parent universe. A typical spherical manifold observed in 
numerical simulations will contain many structures of this kind,
of various sizes and lengths of boundaries, called ``necks''. 
The shortest possible neck has length three, and the associated
baby universe will be called a minimal baby universe (minbu). 
We may also have baby universes characterized by longer 
(but finite and small) neck sizes.

The existence of baby universes opens a completely new possibility for
updating the matter sector of the theory. Note that for a finite 
boundary length $l$, the interaction between the matter in the baby universe
and the parent universe will be of order $l$. 
Take the example of Ising spins: even if we flip 
all spins inside the baby universe, the change in the action will 
only depend on the interactions across the boundary, and not the size $n$
of the baby universe. 
This is a completely different situation than on a
regular lattice where large domains always have large boundaries. 
We may use this observation to define a matter
update which induces a large change in magnetization and has a
large acceptance. 
The efficiency of such an algorithm will depend on the typical
baby universe size.
One possible algorithm consists in searching the triangulation for
a minimal neck and flip all spins inside the associated minbu
with probability $W$.
The same technique can be applied to massless scalar
fields. Here we can perform two operations on the fields in the minbu:
change all of their signs and/or add the same constant to all of them.

The existence of minbus can be exploited also in the geometric sector.
For example, we may use the following three-step algorithm.
\begin{itemize}
\item[$\bullet$] Locate a minimal neck in the triangulation, 
and cut open the original spherical manifold along the neck. 
Both of the resulting 
discs are of the form of a triangulated sphere with a 
triangle removed. 
\item[$\bullet$] Close off both holes by a single triangle, 
to obtain two spherical triangulations of volumes $n+1$ and $N-n+1$ 
(if the original triangulation had volume $N$).
\item[$\bullet$] Remove an arbitrary triangle from each of the two manifolds, 
and glue the resulting discs along their triangular boundaries.
\end{itemize}

We must compute the correct probability factors for such a move and
eventually also include changes in the matter fields.
Based on this idea, an extremely efficient ``minbu surgery'' algorithm  
has been constructed, shortening the autocorrelation
times by three orders of magnitude \cite{ms}. 
The move does not even require 
a lot of geometric updating if it is accepted, 
since the connectivities of the manifold are
changed only locally. 
The minbu surgery usually supplements the
local moves described earlier, although in 2d one may
construct an algorithm that is based exclusively
on the cutting and pasting of baby universes,
if one permits also longer neck lengths.
(A flip move may be viewed as a particular realization
of this, if instead of minbus we consider baby universes with
boundaries of length $l=4$.)

\subsection{Observables in 2d Euclidean gravity}\label{observables}

The Monte Carlo method described above can be used to 
generate a set of uncorrelated configurations of volume $N_2$
which are distributed
according to their Boltzmann weights $e^{-S}$.  
In order to understand the finite-size effects,
one needs to perform the simulations at different values of $N_2$.
The sizes that have been used in numerical
simulations to date range from 1000 to 128000 simplices. 

In 2d it speeds up the calculations to extend the class of 
allowed triangulations
to include also loops of link length one and two (corresponding to
self-energy and tadpole subdiagrams on the dual lattice). 
Let us denote this extended class by $\cT$ 
and the corresponding class of labelled triangulations
by $\tilde{\cT}$. Although the complexes constructed in this way are no
longer simplicial manifolds, 
we can still keep the notions of global topology, of local neighbourhoods
and of a geodesic distance. Likewise,
the relations between the numbers $N_k$ remain unchanged. 
In some cases (including pure 2d gravity), models
based on this set of geometries
can be compared directly with the analytic solutions of corresponding
matrix models, where the exclusion of tadpole and self-energy subdiagrams
corresponds merely to a finite renormalization of the bare coupling constants. 
On $\cT$, the numerical simulations become
even simpler.
For fixed volume, the flip move is still ergodic and also the
minbu surgery moves can be generalized.
The advantage of using this class of triangulations is
a reduction of the finite-size effects, since it turns out that
the local restrictions on the connectivity
do not affect the scaling properties of the system.

Our next step will be to describe the measurement of suitable
``observables'' on the ensemble of configurations generated by
the Monte Carlo algorithm.
The observables most easily obtained are the critical exponents
related to the geometry or the matter fields.
In two dimensions, one can sometimes obtain such observables
analytically, and use them to test the validity of the
numerical results.
Recall that in 2d we start from the partition function
\beq\label{part2}
Z = \sum_{\tilde{\cT}} {1\over N_0!} \sum_{\phi_i} 
e^{-\mu V-S_{matter}(\phi_i)} = \sum_V e^{-\m V} Z_V,
\eeq
where $Z_V$ is the partition function at fixed volume.
If the central charge of matter is $c < 1$, it can be
shown analytically that $Z_V$ behaves like
\beq\label{entropy}
Z_V \sim e^{\m_0 V} V^{\gamma_{str}-3}(1 + \cO(1/V))
\eeq
for large $V$ and spherical topology.
The subleading power
contains the critical exponent $\gamma_{str}$, which has a known
dependence on $c$ (see formula \rf{KPZ} below).
The pure-gravity case corresponds to $\gamma_{str}=-1/2$.

There are other statistical systems 
whose partition function behaves like \rf{entropy}, most notably,
various realizations of branched polymers \cite{baby,branched}. 
In these models,
$\gamma_{str}$ is positive and $\leq 1/2$,
and $V$ counts the number of vertices.

Measuring the distribution of minbu sizes for a triangulation of
fixed volume $N_2=V$ is an efficient device for determining
the critical exponent $\gamma_{str}$ \cite{jain1,jain2}. 
The average number $\la b(n)\ra_V$ of minbus with volume $0 \ll n < V/2 $ 
is given by 
\beq\label{minbu}
\la b(n)\ra_V \sim \frac{n Z_{n} (V-n) Z_{V-n}}{Z_V},
\eeq
since a minbu of size $n$ can be regarded as a spherical triangulation
of volume $n+1$ with one marked triangle. (Since both $n$ and $V-n$ are
assumed large we can neglect small corrections to the volume.) 
Using \rf{entropy}, we obtain
\beq\label{minbu1}
\la b(n)\ra_V \sim V\left( n (V-n) \right)^{\gamma_{str}-2}.
\eeq

Measuring $\gamma_{str}$ gives us information about 
the fractal structure of the theory and provides a
simple test of algorithms in cases where its value is known. 
In all cases with $c < 1$, one finds an excellent numerical
agreement with the predicted values. 
From relation \rf{minbu1} we see that for 
$-1 < \gamma_{str} < 0$, the average minbu size $\la n \ra_V$ remains finite,
but that $\la n^2 \ra_V \sim V^{\gamma_{str}+1}$. 
This means that with growing
$\gamma_{str}$ we will observe increasingly large fluctuations 
in minbu size. If $\gamma_{str} > 0$
the average minbu size behaves like $\la n \ra_V\sim V^{\gamma_{str}}$. 

Another important observable that can be studied 
by numerical simulations is the volume-volume correlator,
which is a particular example of a geometric
two-point function. In the continuum theory,  
it can be defined as
\beq\label{1.1}
G_\L(R) = \int \cD[g_{\mu\nu}]\cD \phi \;e^{-S[g,\phi]} 
\int\!\! \int \! \d\xi\d\xi' 
\sqrt{g(\xi)}\sqrt{g(\xi')} \; \del(d_g(\xi,\xi')-R),
\eeq
where $d_g(\xi,\xi')$ denotes the geodesic distance between
two points $\xi$ and $\xi'$, calculated with respect to
the metric $g_{\mu\nu}$. $G_\L(R)$ is the partition
function for universes with two marked points separated by a geodesic distance
$R$, which we already encountered in Section \ref{fract}. 
In a discretized theory we can construct the analogous quantity
\beq\label{1..discr}
G_\m(r) = \sum_{\tilde{\cT}} \sum_{\phi_i} e^{-\m N_2 - S_{matter}(\phi_i)}
\sum_{i,j} \delta_{D(i,j),r},
\eeq
where $D(i,j)$ is now one of the integer-valued geodesic distances
introduced at the end of Section \ref{dyna}. 
The two possibilities will in general differ by a finite scaling factor.
For large $r$, one can show that $G_\m(r)$ falls off exponentially like
\beq\label{exp}
G_\m(r) \sim e^{-m(\m)r},~~~~~~r \gg 1/m(\m).
\eeq
For small $r$ we expect
\beq\label{expsmall}
G_\m(r) \sim r^{1-\eta},
\eeq
where $\eta$ is the anomalous dimension of the two-point function.
For $\m \to \m_0$ we make the ansatz $m(\m) \sim (\m-\m_0)^{1/d_h}$
where $d_h$ is another critical exponent of the theory. 
From the functional form of the partition function $G_\mu (r)$,
one can derive the estimates 
$\la V \ra_\m \sim 1/(\m -\m_0)$ and $\la r \ra_\m \sim 
1/(\m -\m_0)^{1/d_h}$. From these one obtains a relation 
between the average linear extension and the average volume of
the configuration, namely,
\beq\label{scale}
\la V \ra_\m \sim \la r \ra_\m^{d_h}.
\eeq
The exponent $d_h$ is called the cosmological Hausdorff dimension.
It is a large-scale property of the average ``quantum geometry'' of the
ensemble, and therefore need not coincide with the dimension 
of the individual triangulations.

The function $G_\m(r)$ has been calculated analytically 
for pure gravity, as explained in Section \ref{fract},
leading to $d_h=4$. As we have already pointed out,
it is convenient in numerical simulations to work with triangulations
of fixed total volume. The two-point functions at fixed and
unrestricted volumes are related by a discrete Laplace transform. 
From definition \rf{1..discr} we have
\beq\label{2..discr}
G_\m(r) = \sum_V e^{-\m V} G_V(r),
\eeq 
where $G_V(r)$ is the partition function at fixed volume $V$.
For a particular configuration of the system, $G_V(r)$ is
measured as follows.
Start with a (dual) vertex $i$ and find all (dual)
vertices at (dual) distance 1 from it. By iterating this process,
the triangulation is decomposed into shells characterized by
their distance $r$ from the initial point $i$. 
Note that the shells will in general be disconnected. 
We can measure the total length of each shell, that is, the
number of vertices in the shell.\footnote{Note that one 
could also measure the number of disconnected parts 
and/or the higher moments of their length distribution.}
This construction is repeated for all starting points and all configurations
in the sample to obtain an averaged distribution.  
The resulting quantity $\la n(r)\ra_V$ is -- up to
a normalization -- a numerical estimate of $G_V(r)$.
It is convenient to use the normalization condition
\beq\label{normx}
\sum_r G_V(r) = V,
\eeq
which leads to an interpretation of $G_V(r)$ as the average 
volume of a shell with radius $r$ for a triangulation of volume $V$. 
For large $V$ one expects that
\beq\label{scaled}
G_V(r) \sim V^{1-1/d_h} F(x),
\eeq
where $x= r/V^{1/d_h}$ is the scaling variable. In terms of $x$,
we have a normalization condition
\beq\label{norm1}
\int dx\ F(x) = 1.
\eeq
From an inverse Laplace transform of relation \rf{exp},
we can deduce that for large $x$, 
$F(x)$ must behave like
\beq\label{largex}
\log F(x) \sim x^{\frac{d_h}{d_h-1}}~~~~~~ x \gg 1.
\eeq
On the other hand, we expect for small $x$ that
\beq\label{smallx}
F(x) \sim x^{d_H-1},
\eeq
where $d_H$ is another Hausdorff dimension characterizing the 
quantum geometry at short distances, and is
related to the anomalous dimension $\eta$.
Measurements of $G_V(r)$ have been performed
for systems with a variety of matter types. 
A simple way to determine the cosmological Hausdorff dimension $d_h$ is 
to plot $G_V(r)$ against $r$. Even if the function $F(x)$ is not known
explicitly, one can obtain an estimate of $d_h$ by
comparing the maxima of these curves for different volumes $V$.
One then plots the rescaled quantities $G_V(r)/V^{1-1/d_h}$ as 
a function of $x$, where according to \rf{scaled} the curves for
different volumes $V$ should fall on top of each other. This 
procedure can be viewed as a {\it finite-size scaling method},
of the type used in the study of critical phenomena in statistical 
physics. In the context of quantum gravity,
it was introduced in refs.\
\cite{syracuse,ajw} for $d=2$, 
and in refs.\ \cite{smit,aj1} for $d=4$.

\begin{figure}
\centerline{\hbox{\psfig{figure=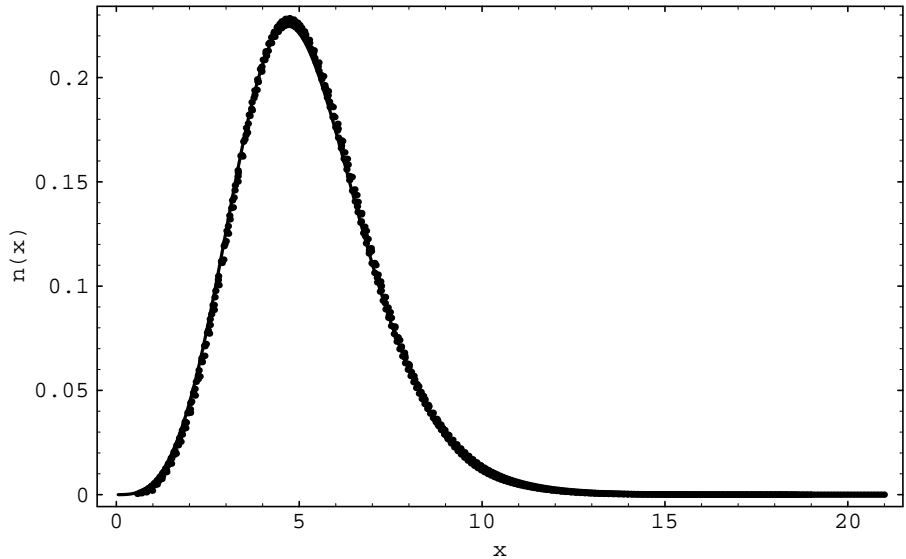,height=9cm}}}
\caption[dh4pure]{The distributions  $n(x)\equiv G_V(x)/V^{1-1/d_h}$
in 2d Euclidean gravity, 
for volumes $V= 1000$, 2000, 4000, 8000, 16000 and 32000, together
with the theoretical distribution $F(x)$.}
\label{dh4pure}
\end{figure}

For pure two-dimensional gravity one can compute $F(x)$ exactly, 
and finds $d_H=d_h=4$. Comparing this to the numerical analysis, it
was discovered that a scaling form with
the correct value of $d_h$ can only be obtained by using a {\it shifted}
scaling variable defined by
\beq
x_{\delta} = \frac{r+\delta}{V^{1/4}},
\eeq
where the constant $\delta$ is determined numerically.
In the case of pure gravity the shift $\delta$ can be 
calculated analytically, but for more complicated systems 
this is usually not the case.
The agreement between theory and numerical simulations 
is impressive. In Fig.\ \ref{dh4pure} we show the 
scaling of $G_V(r)$ for various volumes $V$ {\it and} the 
theoretical curve (which can hardly be seen since it 
coincides perfectly with the numerical data).

Such a shift is also necessary in more complicated systems.
It was soon realized that for the case of pure gravity
this shift is a leading-order finite-size correction to the
scaling form \rf{scaled}, which was obtained for $V \to \infty$.
This improved scaling could also be used to check the small-$x$
behaviour of $F(x)$ with very good accuracy. 

\begin{figure}
\centerline{{\epsfxsize=2.5in \epsfysize=1.67in \epsfbox{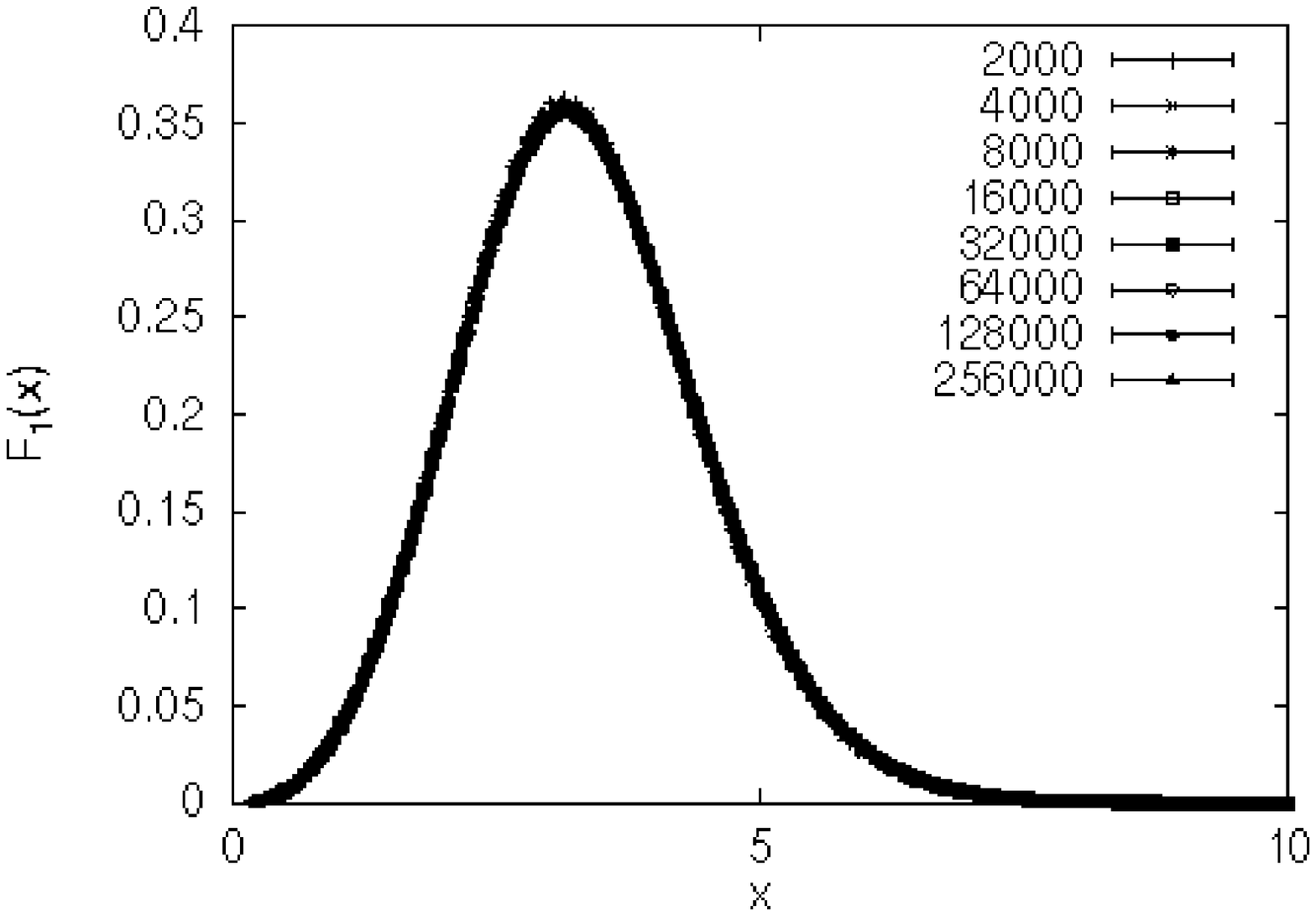}}
{\epsfxsize=2.5in \epsfysize=1.67in\epsfbox{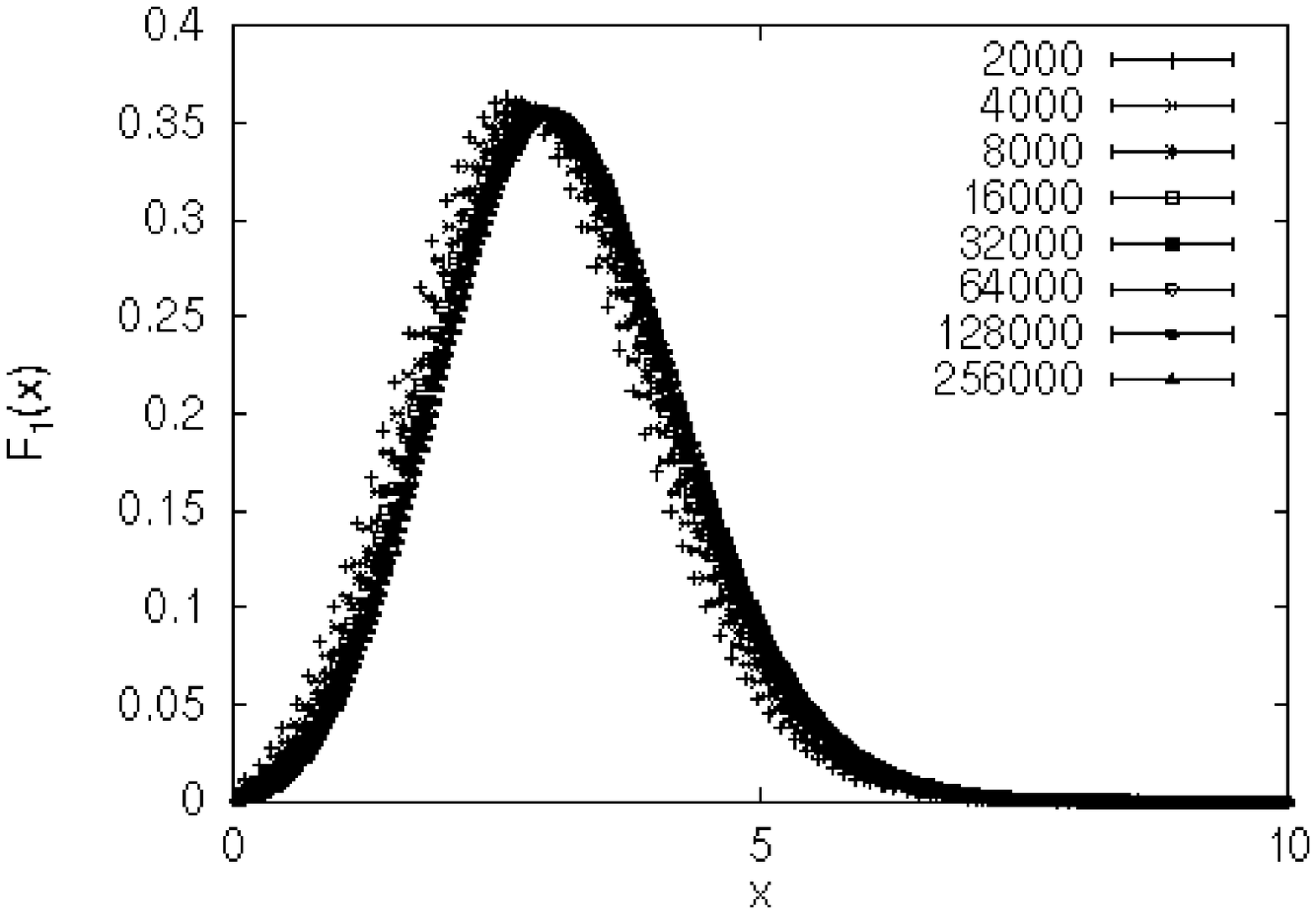}}}
\centerline{{\epsfxsize=2.5in \epsfysize=1.67in \epsfbox{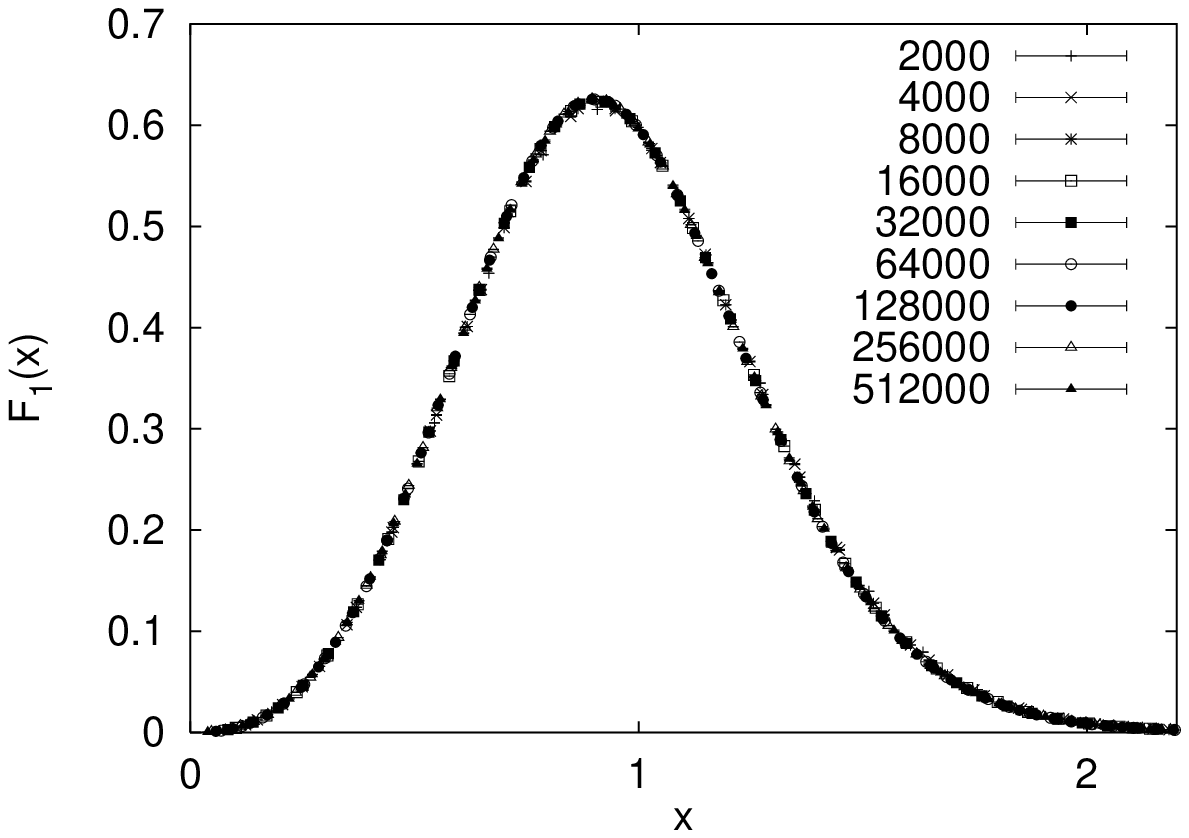}}
{\epsfxsize=2.5in \epsfysize=1.67in \epsfbox{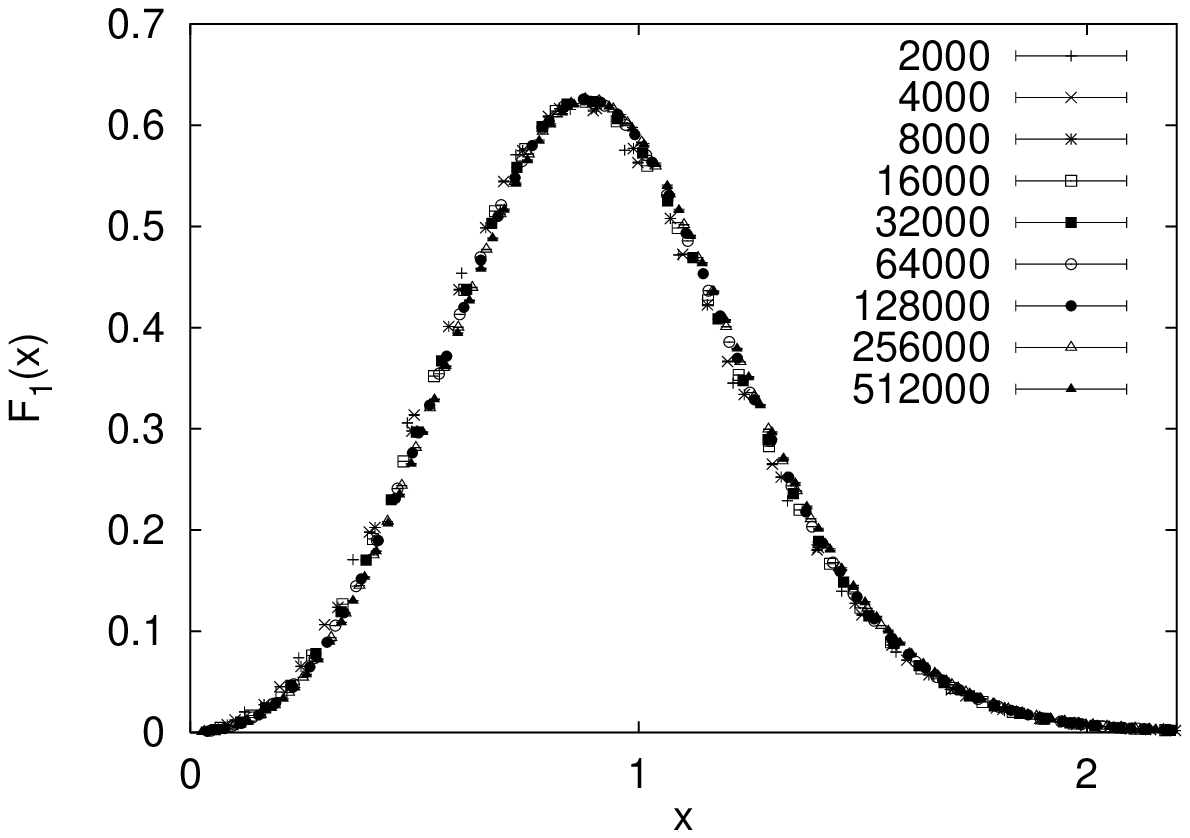}}}
\caption[d-2data]{The two upper figures depict the measured 
distributions $G_V(r)$ of 2d Euclidean gravity coupled to 
$c=-2$ conformal matter, defined in terms
of the dual link distance and rescaled according to eq.\ \rf{scaled} 
to get $F(x)$,
with a shift $\delta = 4.5$ (left) and without a shift (right).
In the two lower figures, the link distance has been used,
again with a shift 
$\delta =0.5$ (left) and without(right). The best fit for the 
Hausdorff dimension 
$d_h$ extracted from the data is $3.58\pm 0.01$, in agreement with 
theoretical predictions \cite{watabiki1}.}
\label{d-2data}
\end{figure}

Another important test of our numerical methods 
is to check whether the scaling limit of the two-point function
is indeed independent of the detailed definition of 
the distance function. Numerically it is
easy to measure the two-point function $\bar{G}_V(r)$,
where $r$ is now the {\it link} distance on the triangulation.
In this case we have no exact analytical prediction, but we may
again fit the numerical results (which look very similar to those
for $G_V(r)$) to the scaling relation \rf{scaled}. 
After a trivial rescaling of the
distance, the matching with the theoretical curve $F(x)$ is even better 
than previously, but again one must include a shift $\bar{\delta}$. 
Numerically, the shift is much smaller than before. This was to be
expected, since a shift in $x$ compensates for small-distance 
artifacts, which are larger on a lattice with smaller vertex 
valence. The aspects mentioned above are illustrated in Fig.\ \ref{d-2data}
by data obtained in the study of a $c=-2$ conformal field theory 
coupled to 2d Euclidean gravity \cite{d-2}. 

Analogous results hold for branched-polymer systems, whose
two-point functions can be calculated exactly.
The distance $r$ is in this case given by the (unique) number of
links separating each pair of points. 
For fixed volume $V \to \infty$, we find a universal scaling of
the form \rf{scaled}, but now with an (intrinsic) Hausdorff
dimension $d_h=2$, which also coincides with the value for $d_H$.
Again a shift in $x$ is needed as a leading finite-size correction 
to obtain optimal agreement with the theoretical results. 

To summarize, measuring $G_V(r)$ or $\bar{G}_V(r)$ provides
an efficient way for determining the Hausdorff dimension $d_h$,
also in cases where the two-point functions are not known exactly. 
To take care of short-distance effects, it is important to work with
an improved, shifted scaling variable $x_{\delta}$.
In practice one measures $G_V(r)$ for a number of volumes $V$, 
and tries to find
a best fit to the universal scaling relation \rf{scaled} 
by fitting both $\delta$ and $d_h$. 

\begin{figure}
\centerline{\hbox{\psfig{figure=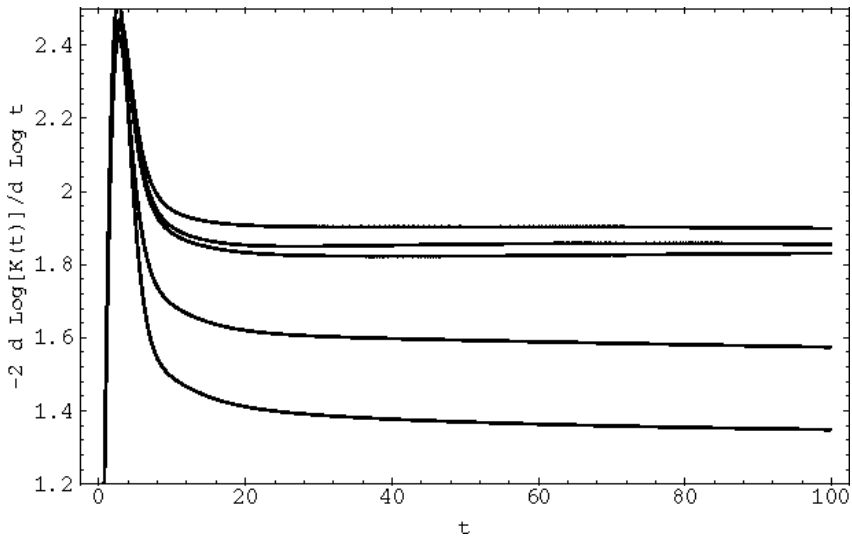,height=7cm}}}
\caption[spek]{The spectral dimension $d_{s}\approx 
-2 d\log P (T)/d \log T$
versus $T$ for  $c=0$ (top curve), 
$c=1/2$, $c=1$, $c=3$ and c=5 (bottom curve)
theories coupled to 2d Euclidean quantum gravity. The system size 
is $V$=16000 triangles.}
\label{spek}
\end{figure}

Another geometric observable characterizing a compact metric manifold
is its {\it spectral dimension}, which is 
related to the diffusion equation and
the spectrum of the Laplace operator.
If a diffusion process is started with a completely localized initial
condition, one can measure the return probability $P(T)$ of a
fictitious test particle after ``time'' $T$. 
On a fixed continuum geometry, one obtains after averaging over
the initial point that
\beq\label{spect}
P(T) \sim {1\over T^{d_s/2}}\sum_k a_k T^k,
\eeq
for small $T$, where the coefficients $a_k$ can be expressed 
in terms of local geometric invariants.
We can measure $d_{s}$ also in the discretized theory, 
but some care must be exercised when comparing with a
continuum formula like \rf{spect}, since it is exactly the short-time 
limit $T \to 0$ that is ill-defined in the discrete case.
This can already be seen 
in the simple case of diffusion on a discrete one-dimensional
line, with a discrete Laplacian and a continuous time. 
The diffusion equation
\beq\label{diffusion}
\frac{d\phi_i}{dT} = \frac{\phi_{i+1}+\phi_{i-1}-2\phi_i}{2},
\eeq
with the initial condition $\phi_i(0)=\delta_{i0}$
can be solved exactly, yielding 
\beq\label{example}
P(T) = e^{-T} I_0(T) \sim {1\over T^{1/2}}(1 +\cO(1/T)),
\eeq
where $I_0(T)$ is the Bessel function. One therefore 
rederives the correct spectral
dimension ($d_s=1$) only for {\it large} $T$. 
The short-time behaviour is completely different and
dominated by discretization artifacts. 
For this simple example, one can check
numerically the effect of the finite system size, by
changing the discrete line to a closed circle with $V$ points.
One observes three distinct regions:
\begin{itemize}
\item[$\bullet$] small time $T$, which is similar to the example above, 
and dominated by discretization effects;
\item[$\bullet$] intermediate $T$, where one obtains the correct value of 
$d_s$; and
\item[$\bullet$] very large $T$, where the system approaches a 
stationary translation-invariant state.
\end{itemize}
A similar structure is also expected in less trivial cases, such as
simplicial quantum gravity in two dimensions and higher.  
This behaviour is illustrated in Fig.\ \ref{spek}, where we show
the results of measuring the spectral 
dimension for 2d Euclidean quantum gravity coupled to matter fields 
of various central charges $c$.

The measurement of the spectral dimension in 2d Euclidean 
quantum gravity is a nice illustration of the 
fruitful interaction between numerical ``experiment'' and theory.
Since the quantum geometry of 2d Euclidean gravity has a fractal 
Hausdorff dimension $d_{h}\equ 4$, one could also have expected an anomalous
spectral dimension $d_{s}$. 
However, numerical simulations consistently found that 
$d_s =2$, no matter which field theory was coupled to 
gravity, as long as the central charge $c$ remained $\leq 1$.
This inspired an analytic proof of this value, using continuum 
conformal Liouville theory  and an assumption about  finite-size 
scaling \cite{abnr,alnr}. We should point out that $d_s \equ 2$ 
does not imply that one cannot see the fractal structure 
corresponding to $d_h \equ 4$ in the diffusion process.
It manifests itself in the relation 
\beq\label{diffrel}
\langle R \rangle_T \sim T^{1/d_h}
\eeq
between the ``time'' $T$ and the average 
geodesic distance $R$ diffused in that time. We refer again
to \cite{abnr,alnr}
for details.

For $c > 1$, the numerical estimate was close to $d_s=4/3$.
Eventually it was proven that 
generic branched polymers have a spectral dimension $d_s = 4/3$ 
\cite{john},  thus reconciling theory and numerical results. 
At the same time, it 
provided independent evidence that for $c > 1$ the space-time 
degenerates into branched polymers.

We have so far discussed only purely geometric observables. 
For theories coupled to matter, there will be additional
critical exponents characterizing the behaviour of the system close 
to the critical matter coupling $\beta_c$.
Near this transition point, we expect to find a divergent 
correlation length $\xi(\beta)
\sim |\beta-\beta_c|^{-\n}$, for some positive $\n$.  
Since we always study systems of finite size, 
we will never observe a genuine
phase transition, but merely a pseudo-critical point $\beta_c^*(V)$,
where the linear extension $L$ of the system behaves like $L \sim \xi(\beta)$.
For a finite system $L^d = V$.
A system with a dynamical geometry may possess two scales:
one associated with the cosmological constant when $\m \to \m_0$ and another
one related to the matter phase transition. Using standard finite-size
scaling arguments, we expect that the measured values 
of the specific heat $c_v$, the magnetization per unit volume $m$,
and the magnetic susceptibility $\chi$ at the pseudo-critical point 
$\beta_c^*$ behave like
\bea
c_v(\beta_c^*) &\sim& V^{\frac{\alpha}{\n d}}, \nonumber\\
m(\beta_c^*) &\sim& V^{-\frac{\beta_m}{\n d}}, \nonumber\\
\chi(\beta_c^*) &\sim& V^{\frac{\gamma_m}{\n d}}, \label{criticl}
\eea
where $\alpha$, $\beta_m$ and $\gamma_m$ satisfy the scaling relations
\beq\label{relations}
\alpha + 2\beta_m +\gamma_m =2,~~~~~~2\beta_m+\gamma_m=\n d.
\eeq
For magnetic systems corresponding to $0 < c < 1$, one 
knows from analytical calculations 
the values of the critical exponents as well as the 
values of the critical points $\beta_c$. 

A relation between the scalings in the geometric and matter sectors can be
obtained by generalizing the two-point functions $G_\m(r)$
and $G_V(r)$, to include also correlations between the matter fields at 
geodesic distance $r$. One example of a quantity of this kind is
\beq\label{3..discr}
G_\m^{\{\phi,\psi\}}(r) = \sum_{\tilde{\cT}} \sum_{\phi_i} 
e^{-\m N_2 - S_{matter}(\phi_i)}
\sum_{i,j} \phi_i \psi_j\delta_{D(i,j),r},
\eeq
which may be expressed as the discrete Laplace transform 
of a finite-volume correlator,
\beq\label{4..disrc}
G_\m^{\{\phi,\psi\}}(r) = \sum_V e^{-\m V} G_V^{\{\phi,\psi\}}(r).
\eeq 
In this expression, $\phi_i$ and $\psi_j$ are local functions
of the matter or geometric variables at the points $i$ and $j$. 
A typical example is the spin-spin
correlation function of the Ising system. At the combined critical 
point of the Ising model and the cosmological constant, 
\rf{3..discr} yields a reparameterization-invariant definition of  
a matter correlator in quantum 
gravity. More precisely, the correlation 
function is that of a $c\equ 1/2$ conformal field theory coupled to
2d Euclidean quantum gravity. Note also that the previously defined
geometric two-point function \rf{1..discr} is a special case of 
\rf{3..discr}, with the unit operator replacing the spin operators.  

\begin{figure}
\centerline{\hbox{\psfig{figure=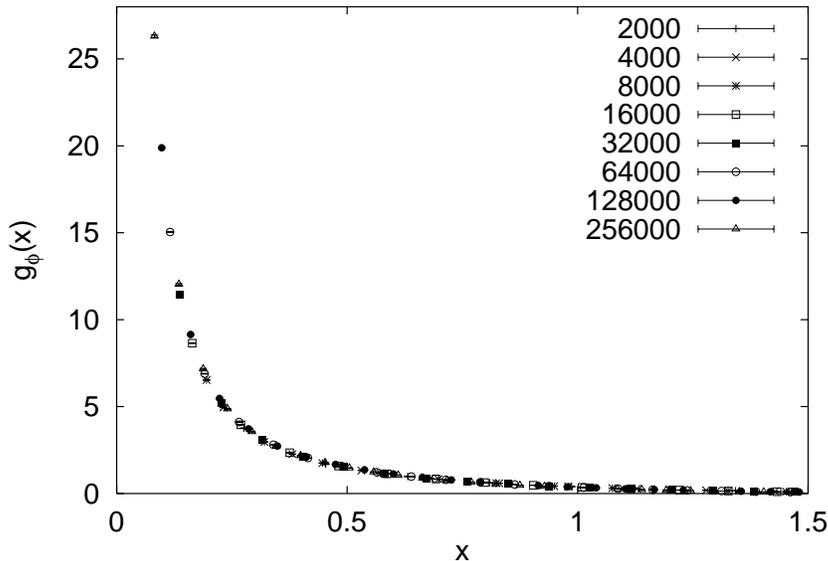,height=8cm,angle=0}}}
\caption[correlator]{The normalized 
spin-spin correlation function for the Ising model coupled to 2d
Euclidean gravity, plotted
as a function of the 
dimensionless length $x=r/N_T^{1/d_h}$, where $r$ is the geodesic 
length and $N_T$ the space-time volume.
The number of triangles $N_T$ ranges from 2000 to 256000 and 
the Hausdorff dimension is $d_h=4.0$.
The short-distance behaviour is in accordance with KPZ scaling.}
\label{correlator}
\end{figure}

This construction has provided us with a
new understanding of the  KPZ-exponents of conformal field theories
coupled to gravity. These exponents may be viewed as the ``dressed''
scaling exponents of the primary fields in the conformal theory.
However, before it was realized that they should be analyzed in
terms of the definition \rf{3..discr}, it was not clear 
how to generalize the usual flat-space correlators because of
the requirement of reparametrization-invariance (see \cite{aa,aa1} for 
further discussion). 
Fig.\ \ref{correlator} shows  
the result of a numerical simulation of the (suitably normalized)
spin-spin correlator.

We should point out a subtlety in the definition of the correlator 
\rf{3..discr}.
If the expectation values of the fields entering are not zero,
one might be interested in defining the corresponding 
{\it connected correlator}. However, there is no straightforward
way of defining such an object since 
\rf{3..discr}, in addition to the usual average over 
field configurations, also includes an average over geometry.
A discussion of possible definitions of the connected 
part of the two-point correlators and their scaling properties
can be found in \cite{ABJ}.

All of the observables introduced above 
can also be used in numerical simulations of higher-dimensional
gravity-matter systems, and -- apart 
from the ones involving baby-universe counting -- 
also for Lorentzian quantum gravity.

\subsection{Comments on the 2d results}\label{results}

Our obvious starting point in 2d Euclidean 
quantum gravity was the measurement of $\gamma_{str}$, in order to
compare it with the
theoretical prediction for $c<1$, namely,  
\beq\label{KPZ}
\gamma_{str}=\frac{c-1-\sqrt{(c-1)(c-25)}}{12}.
\eeq
For $c>1$, this quantity becomes imaginary and its interpretation 
in terms of Liouville theory breaks down. For $c=1$, 
logarithmic corrections appear. 
For a number of special values of $c$, we have explicit discrete
models, which at their critical points represent 
conformal field theories with charge $c$. They are
pure 2d gravity ($c=0$, $\gamma_{str}=-1/2$),
the Ising model ($c=1/2,~\gamma_{str}=-1/3$),
the 3-state Potts model  ($c=4/5,~\gamma_{str}=-1/5$),  
the 4-state Potts model  ($c=1,~\gamma_{str}=0$), 
and a single massless scalar field ($c=1,~\gamma_{str}=0$). 
In all of these cases, formula \rf{KPZ}
has been verified with great accuracy. 
A coupling of several matter fields corresponds to adding their central
charges.

\begin{figure}
\centerline{\hbox{\psfig{figure=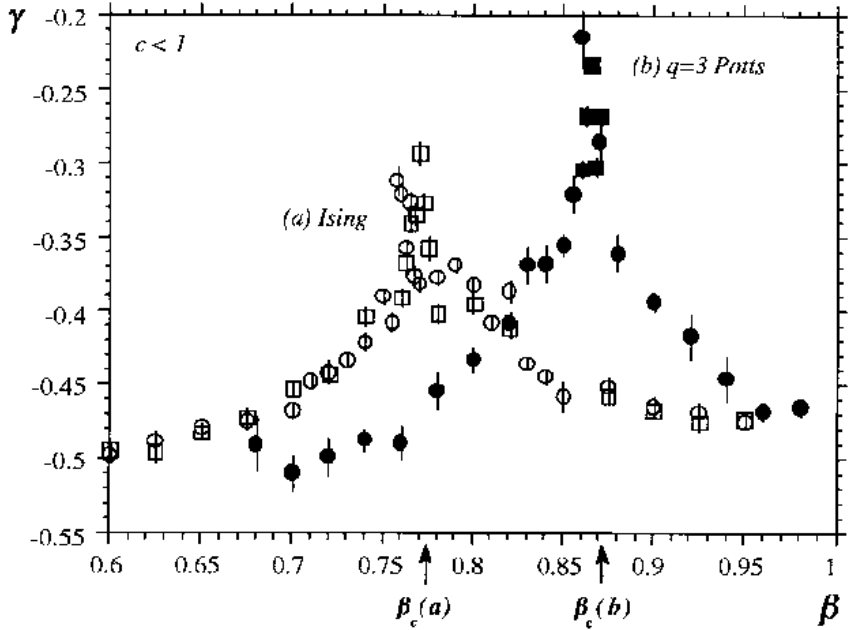,height=9cm}}}
\caption[gamstr]{Measurements of $\g_{str}$ as a function of $\beta$, 
for the Ising model (white),
 and the $3$-state Potts model (black), at system sizes 
$N$ = 1000 (circles) and 2000 (squares).}
\label{gamstr}
\end{figure}

Away from their critical
points $\beta_{c}$, the finite spin systems approach a pure-gravity
behaviour, with $\gamma_{str}\to -1/2$.
This observation can be turned into a method for locating $\beta_{c}$,
by monitoring the change in $\gamma_{str}$ as the coupling $\beta$
is varied (Fig.\ \ref{gamstr}).

For all the spin systems, measurements of 
the magnetic critical exponents defined in \rf{criticl} are
in perfect agreement with the theoretical predictions\footnote{This 
also includes the case $c=1$, which can be realized 
either as two Ising spins, as a $q\equ 4$ Potts model, or as 
a single Gaussian field coupled to gravity. However,
there are logarithmic corrections
to formula \rf{entropy} which have to be included in the fits in order 
to obtain $\g_{str}\equ 0 $ from the data.}. 
We conclude that for the matter models with  $0 \le c \le 1$
which have been used in numerical simulations, 
the scaling limit of the discretized theory
corresponds to a unitary matter model coupled to Liouville gravity.

A somewhat surprising result in this range of $c$ is the numerical 
value of $d_h$ obtained from the scaling of $G_V(r)$.
For $0\le c < 1$, there are two theoretical predictions for $d_h$,
namely  
\beq\label{dh1}
d_h^{(i)} = 2 \frac{\sqrt{25-c}+\sqrt{49-c}}{\sqrt{25-c}+\sqrt{1-c}},
\eeq
derived by studying diffusion in a fluctuating geometry, 
using Liouville theory \cite{watabiki1}, and 
\beq\label{dh2}
d_h^{(ii)} = \frac{24}{\sqrt{1-c}\ (\sqrt{1-c}+\sqrt{25-c})},
\eeq
derived from matrix model considerations \cite{ik1}.
The formulas agree for 
$c\equ 0$ (where $d_h \equ 4$), but differ elsewhere. In both derivations 
there are  plausible, but unjustified, assumptions.
Neither of them could be confirmed by numerical 
simulations\footnote{It is now understood that formula \rf{dh2}
measures the fractal dimension of (generalized) spin clusters 
of the matter fields, rather than the fractal dimension of the 
underlying geometry \cite{ajk}.}.
Instead, a numerical estimate $d_{h}\simeq 4$ was obtained 
for all cases with $0 \le c \le 1$  \cite{ajw,aa}
together with $d_H \simeq d_h$ \cite{aa1,aa} from the small-$x$ behaviour 
of $F(x)$. The numerical errors seem sufficiently small 
to support the conjecture
that $d_h=4$ for unitary matter theories coupled to 2d gravity.
The spin-spin correlation functions measured for the Ising system
($c=1/2$) at $\beta=\beta_c$ seem to indicate that the matter
degrees of freedom do not introduce a new scale and that also
for these correlations, there is a universal scaling corresponding
to $d_h \simeq 4$.

Our numerical information for the case $c < 0$ is only partial.
A special case is $c=-2$, where the simulations can be based
on a direct generation of diagrams, instead of Monte Carlo
methods. As already mentioned in connection with Fig.\ \ref{d-2data} 
this led to $d_h=3.58 \pm 0.02$, which is very close to the value of
$d_h=3.561\dots$ predicted by \rf{dh1}. Less conclusive numerical 
data for $c=-5$ also show agreement with \rf{dh1}.
Formula \rf{dh1} also predicts that $d_h=2$ as $c \to -\infty$, which is 
the expected flat-space behaviour. It thus seems that eq.\ \rf{dh1}  
could be correct for $c \le 0$. It is still not understood 
why the formula fails for $0 < c \le 1$.

For $c>1$, the minbu size diverges as $V \to \infty$. Systems with
$c > 1$ can easily be studied numerically by coupling several copies of 
spins or scalar fields. In all cases studied, $\gamma_{str}$ was found
to be positive and to approach $+1/2$ for $c \simeq 4$ and bigger, 
suggestive of a branched-polymer system. 
This interpretation would also agree with
the scaling analysis of $G_V(r)$ which is consistent with $d_h \to 2$,
and the measurement of $d_s$, consistent with $d_s \to 4/3$.
It is not clear to what extent the observed behaviour is universal,
since the numerical estimates depend on the details of the regularization.
For example, the approach to the branched-polymer phase seems slower 
when regular triangulations $T$ rather
than generalized configurations in $\cT$ are used. 
This could be due to finite-size
effects, or alternatively to the presence of other universality classes
of branched-polymer systems. 

In the 2d Euclidean case, we have learned from both theoretical and 
numerical investigations that there is only a small ``window'' of 
physically sensible theories, by which we mean
unitary matter models coupled to gravity. If we move
outside this range by 
adding too many matter fields, the theory breaks down. 
This indicates that matter and geometry are interacting strongly. 

Baby universes play a crucial role in understanding the nature of the
spin-gravity interaction, highlighting at the same time the difference 
between the Euclidean and Lorentzian gravity models.
Again the Ising model serves as an ideal illustration.
The Ising ground state at zero temperature, for both fixed and 
fluctuating geometries, is the state where all spins are aligned.
Since the energy of a given spin configuration is proportional to 
the length of the boundary separating spin-up and spin-down regions, 
the dominant spin configurations at low temperature (equivalently, 
at large spin coupling) are those with minimal spin boundary lengths.

\begin{figure}
\centerline{\input{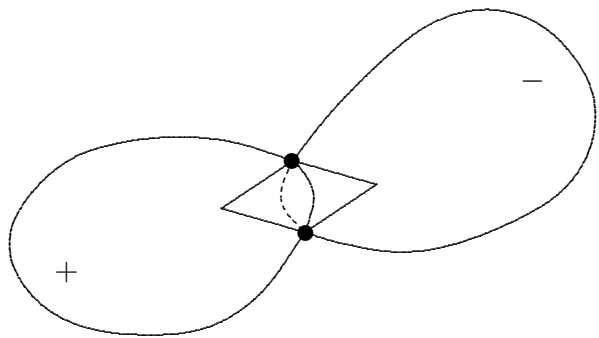}}
\caption[isingbabies]{Two spin clusters separated by only two links
which form a ``bottleneck'' on the surface. Two of the four triangles 
containing one of the two spin-boundary links are also shown.}
\label{isingbabies}
\end{figure}

The lowest spin excitations contributing to the free energy density
come from spin clusters with boundaries of minimal length. 
Unlike on flat, regular lattices, a short spin boundary does not
imply that there are few spins inside. On the contrary, the typical
situation in Euclidean quantum gravity has an entire baby 
universe of up-spins, say, on one side of the minimal boundary,
and the parent universe with opposite spin orientation on the other,
as illustrated in Fig.\ \ref{isingbabies}. 
Since the baby universe can have any size, there is
no restriction on the size of spin clusters, even at low temperature.
The introduction of a fluctuating geometry with baby universes 
thus has a strong effect on the matter behaviour. In turn, since
short spin boundaries are energetically preferred, the matter has
a tendency to ``squeeze off'' the underlying geometry, generating
even more baby universes. 

\begin{figure}
\centerline{\input{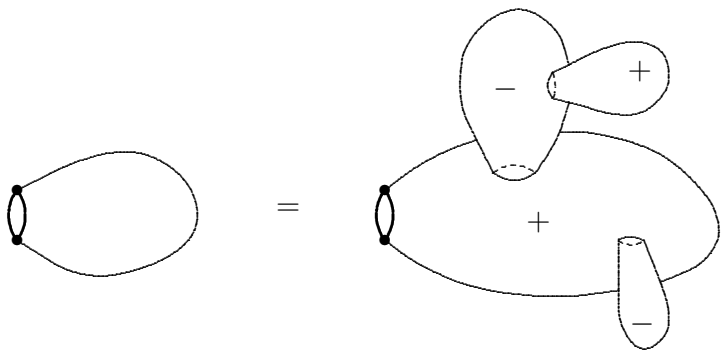}}
\caption[proliferation]{A surface with a minimal boundary and 
its recursive decomposition into baby universes associated with 
spin clusters.}
\label{proliferation}
\end{figure}

When more than two Ising models are coupled to 2d Euclidean quantum gravity,
the matter-geometry interaction becomes so strong 
that the geometries degenerate into so-called branched 
polymers, which can be viewed as infinitely branched trees of 
baby universes of cut-off size (the lattice spacing $a$).
This provides us with an explicit picture of the $c\equ 1$ barrier 
of 2d Euclidean quantum gravity. The proliferation of baby universes
and its relation to spin clusters are illustrated in Fig.\ \ref{proliferation}.

Since baby universes are absent from Lorentzian gravity,
its interaction with matter is much weaker. 
The Ising model on Lorentzian geometries has not yet been
solved analytically, but its critical exponents have
been determined both by a high-temperature expansion and 
Monte Carlo techniques \cite{aal1}.
The Hausdorff dimension of space-time is still $d_{h}=2$,
and the critical Ising exponents retain the 
Onsager values found on fixed, regular lattices, in spite
of large fluctuations of the geometry. It has also been 
shown \cite{lotti} that a particular dimer model\footnote{It is known 
that the critical behaviour of dimer models is associated with
a $c\equ \mi 2$ conformal field theory. Since the dimer model 
considered in \cite{lotti} imposes certain restrictions on
the allowed dimer positions, it strictly speaking has 
not been proven to lie in the same universality class.} coupled to Lorentzian 
gravity does not change the fractal dimension of the geometry.
It is thus tempting to conjecture that the (fractal) dimension 
of space-time remains equal to 2 as long as the central change 
of the matter fields is less than or equal to 1.

\begin{figure}
\centerline{\hbox{\psfig{figure=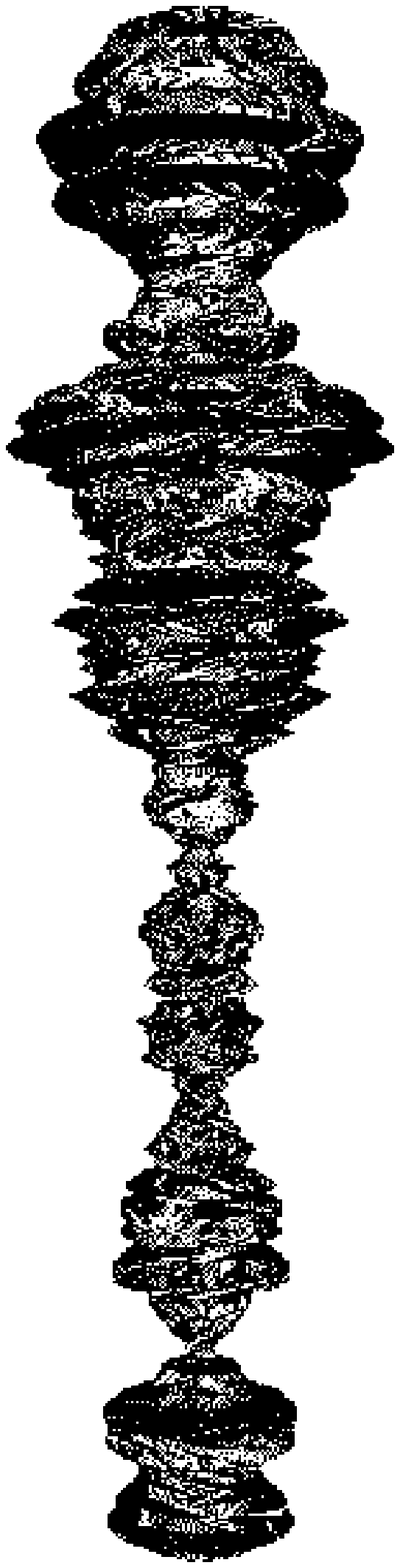,height=8cm,angle=0}}
\hspace{3cm}\hbox{\psfig{figure=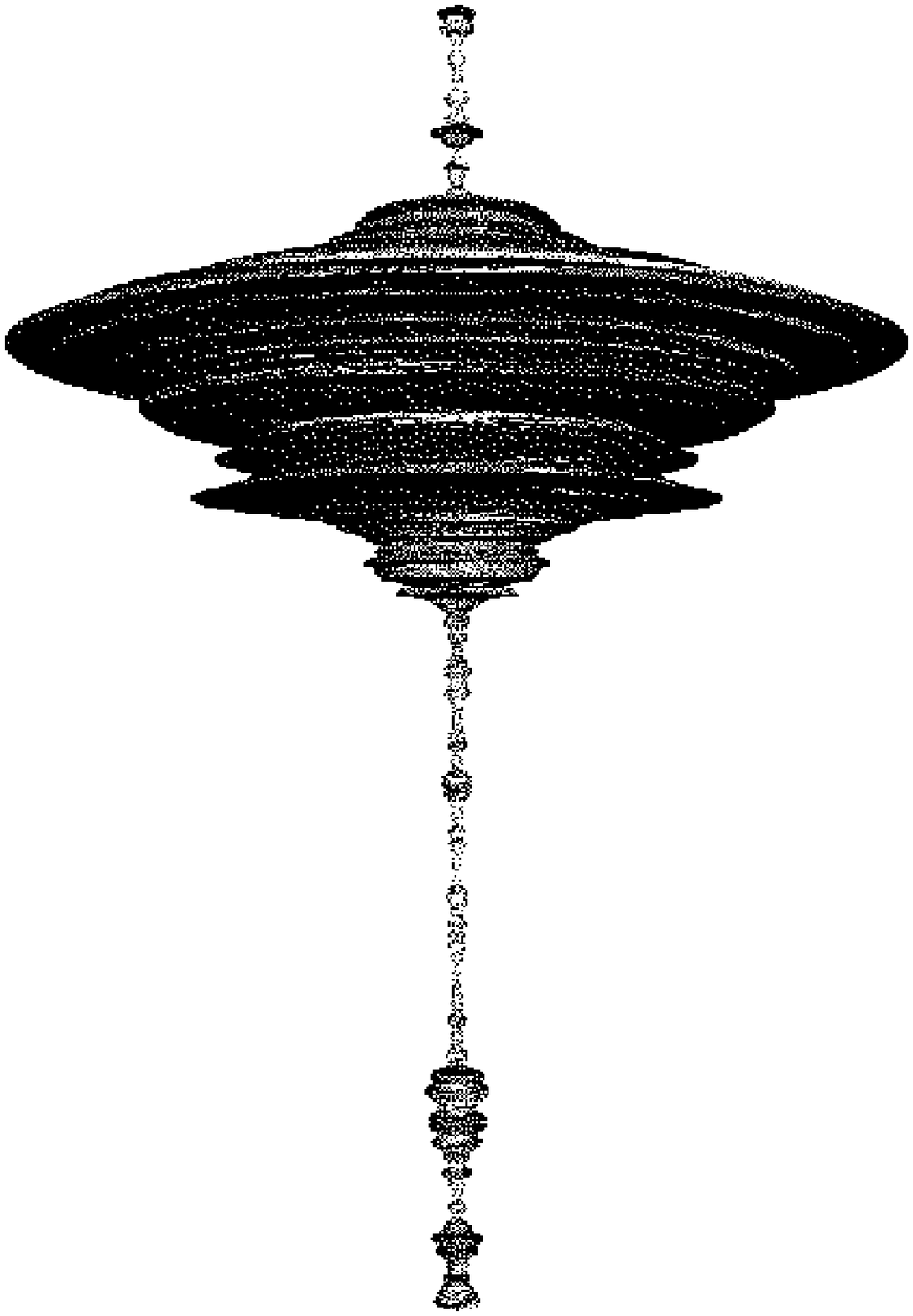,height=8cm,angle=0}}}
\caption[newfig]{Two typical space-time configurations, resulting
from the coupling of one Ising model 
(left) and eight Ising models (right) to Lorentzian gravity.}
\label{newfig}
\end{figure}

It is an interesting question 
whether there is an analogue of the $c\equ 1$ barrier when sufficiently
many matter degrees of freedom (with a sufficiently large
central charge $c$) are added. This is indeed what seems to happen.
A phase transition in the geometry has been observed when
coupling 8 Ising spin copies (corresponding to $c\equ 4$) 
to 2d Lorentzian quantum gravity \cite{aal2}. 
This effect is illustrated in Fig.\ \ref{newfig}, 
where we show typical configurations of
the fluctuating Lorentzian geometry in the case 
of coupling to a single Ising model and to 8 Ising models.
The one-Ising triangulation is qualitatively very similar to the
fluctuating geometry without any Ising spins, 
but in the case of 8 Ising spins, the effect of the matter 
is very pronounced. Since the creation of baby-universe branches
is by definition forbidden, the only way for the matter to create
short spin-boundaries is by squeezing individual spatial slices to
a minimal (cut-off) size, producing the long, stalk-like structure
seen in Fig.\ \ref{newfig}.
In the remainder of space-time, the spatial slices attain
a macroscopic size. On this extended part of the universe,
the scaling of the average spatial length in terms of
the cosmological constant is anomalous, 
\beq\label{mat1}
\la L\ra_\L \sim \frac{1}{\L^{2/3}},
\eeq
which was not the case for $c\leq 1$. Moreover, the critical
Ising exponents, when measured on the extended region, 
are still given by the Onsager values. We therefore have
identified an analogue of the $c=1$ barrier, at which a
geometric phase transition takes place.
However, its effect is much milder than in Euclidean gravity.
Ising-life continues even beyond the barrier. 
This underlines yet again the drastic difference between
2d gravity with and without baby universes.

\section{Dynamically triangulated quantum gravity in $d>2$}\label{higherd}

\subsection{Generalization to higher dimensions}

The general outline of the simplicial regularization of Euclidean quantum
gravity presented above remains true in dimension $d >2$, where the
fundamental building blocks are $d$-simplices with $d+1$ vertices.
The partition function for dynamically triangulated 
pure gravity is given by
\beq\label{highpart}
Z = \sum_{\bar{T}} {1 \over N_0(T) !} e^{-S_{EH}},
\eeq
where the sum is taken over a class of labelled $d$-dimensional 
triangulations with spherical topology. 
In the following, we will be mainly interested in the cases 
$d=3,4$, where the discretized Einstein-Hilbert action \rf{HED} 
can be written equivalently as
\beq\label{SHE}
S_{EH} = \kappa_d N_d - \kappa_0 N_0,
\eeq
using the Dehn-Sommerville relations. 
Since for simplicial {\it manifolds} and fixed volume $N_d$, the number 
of vertices has an upper limit $N_0 \propto N_d/d$, the action \rf{SHE}
is bounded for finite volume.

Note that in $d>2$ it is not possible to realize
a locally flat geometry by gluing together equilateral simplices.
Only in 2d there is a regular tiling of flat space, in which each
vertex is shared by six triangles. This happens because the 
dihedral angle $\theta_d$ of each $d$-simplex contributing to
the parallel transport around a $(d-2)$-simplex 
satisfies $\theta_d = \arccos 1/d$, which 
for $d>2$ is not a rational fraction of $2\pi$. 
However, this is a short-distance property
which should become irrelevant in the scaling limit.

In the simplest case of gravity without matter, there are
two couplings: $\kappa_d$, which is proportional to the  
cosmological constant, and $\kappa_0$, related to the gravitational
constant. 
Both quantities are dimensionless in the lattice formulation,
but will get their dimension back through ``dimensional
transmutation'' in the scaling limit, if it exists.

The gravitational state sum is now
\beq\label{highpart1}
Z = \sum_V e^{-\kappa_d V} Z_V(\kappa_0),
\eeq
where $V=N_d$, and $Z_V(\kappa_0)$ is the partition function
at fixed volume.
In order for the sum \rf{highpart1} to be well defined, $Z_V(\kappa_0)$ must
be exponentially bounded for large $V$,
\beq\label{largev}
Z_V(\kappa_0) \sim e^{\kappa_d^c(\kappa_0) V} z_V(\kappa_0).
\eeq
The existence of such a bound has been shown \cite{carfora,acm}, but
the form of the subleading terms is at present unknown. 
A power behaviour of $z_V$ leads to
a situation similar to \rf{entropy}, but there could in principle 
also be a logarithmic dependence 
$\log z_V \sim V^{\alpha}$, with $\alpha < 1$. For systems with
a fixed volume one can replace $Z_V$ by $z_V$ when averaging over
configurations.
Note that higher-dimensional gravity with the action \rf{SHE} resembles 
a spin system in 2d, with the role of the matter coupling $\beta$ played
by $\kappa_0$. 
The cosmological constant will undergo a renormalization according to
\beq\label{highpart2}
Z = \sum_V e^{-(\kappa_d-\kappa_d^c) V} z_V(\kappa_0),
\eeq
and $1/(\kappa_d-\kappa_d^c)$ gives an estimate of $\la V \ra_{\kappa_d}$.

To understand the critical properties of the system defined by
\rf{highpart2}, one must investigate the limit $V \to \infty$. 
For large positive
$\kappa_0$, configurations with many vertices will be favoured.
Comparing with the 2d case, we expect this limit to correspond
to a branched-polymer phase. Conversely, for large negative 
$\kappa_0$, the system will move towards lower  
$N_0$, leading to ``crumpled'' configurations. 
Note that unlike in 2d, $N_0$ and $V$ are independent, 
and $N_0$ can be very small, for example, it is easy to find 
triangulations such that $N_0 \sim V^a$ with $a < 1$. 
The fraction $p(\kappa_0) = d \la N_0 \ra_V/V$
can be used as an order parameter of the theory, and is related to the average
local curvature. It may vary between 0 and 1, corresponding to the two
extreme geometric phases. In the limit $p \equ 0$ ($\k_0$ small), where  
the geometry is ``crumpled'', there are only a few vertices and one can 
move from one to another in only a few steps. The Hausdorff dimension 
of such geometries is large. In the limit $p \equ 1$ (large $\k_0$), 
the number of vertices for a given $V$ is maximal, 
leading to branched-polymer configurations with Hausdorff 
dimension two. 

Neither of these two limits looks
like a promising candidate for a scaling $d$-dimensional quantum theory.
Let us compare this situation with that of 
a 2d spin system. The two extreme phases correspond to
the high- and low-temperature regions of the spin model,  
where also the spin system has no continuum limit. 
In the high-temperature limit, the spins fluctuate totally 
randomly, whereas in the low-temperature limit they hardly fluctuate at 
all. However, there is a critical temperature $T_c$ 
where the spin fluctuations
become critical and long-ranged. At $T_c$, one can take 
a continuum limit and recover a $c\equ 1/2$ conformal field theory.
In a similar way one might hope to find a critical gravitational
coupling $\kappa_0$, where a transition between the 
two phases of geometries takes place, and where
more ``reasonable'' geometries, exhibiting long-range 
fluctuations, can be observed. 
Such a point could serve as a non-perturbative 
fixed point where a theory of quantum gravity can be defined.

Although the problem of solving higher-dimensional gravity 
with the simple action \rf{SHE} is a combinatorial problem
which superficially looks quite similar to the 2d case, we still 
have no hints of how to find an exact solution. 
There is a mean-field analysis \cite{acm,acm1} which gives a 
surprisingly precise 
description of the phase transition (see also \cite{balls}), but most of 
our understanding of Euclidean dynamical triangulations 
in $d>2$ comes from performing Monte Carlo simulations, along the lines
described in Section \ref{numerical}.
Also in these cases a set of ergodic moves is known \cite{alexander}. 
Recall that in 2d, one possible set was given by the $(p,q)$-moves,
for integer $p,q>0$ and such that $p+q=d+2$.
During a $(p,q)$-move, a subcomplex
of $p$ simplices is replaced by one of $q$ simplices with the same
$(d-1)$-dimensional boundary. This set
can be generalized to higher dimensions \cite{pachner,GV}, using
the observation that by gluing the two subcomplexes from
before and after the move, one obtains the boundary of a 
$(d+1)$-simplex. The boundary can be divided into two connected 
parts (a part $X$ and its complement $\bar X$) in a finite number of ways. 
Each move corresponds to replacing some $X\to\bar X$.
In two dimensions, the boundary is that of a tetrahedron, 
made of four triangles. It can be divided either into two pairs of 
two adjacent triangles or into a single triangle and a set of three
triangles sharing a vertex of order three.

\begin{figure}
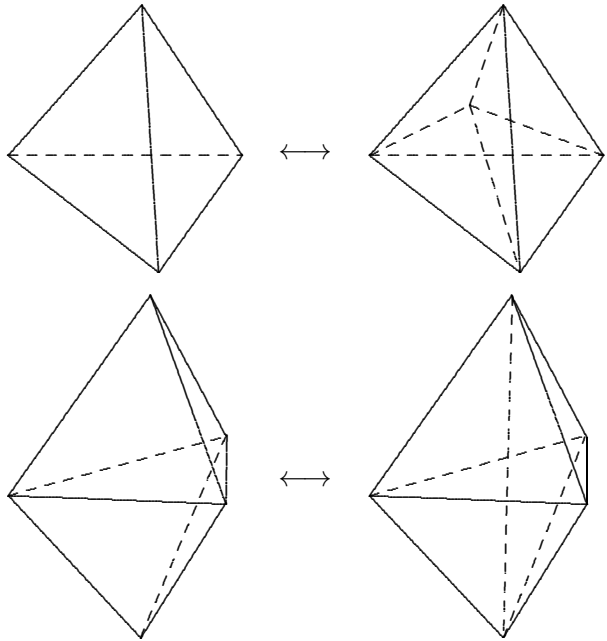

$$
\raisebox{1ex}{\parbox{3.5cm}{\input{fig6-4a}}} \longleftrightarrow ~~~
\raisebox{1ex}{\parbox{3.5cm}{\input{fig6-4b}}}
$$
$$
\raisebox{1ex}{\parbox{3.5cm}{\input{fig6-4c}}} \longleftrightarrow ~~~
\raisebox{1ex}{\parbox{3.5cm}{\input{fig6-4d}}}
$$
\caption[d3move]{The elementary ergodic moves in $d=3$.
There are four different moves since the moves are different from
their inverses.}
\label{d3move}
\end{figure}

Performing an analogous construction in 3d, one obtains 
$(1,4)$- and $(2,3)$-moves, together with their inverses $(4,1)$ 
and $(3,2)$. The first move inserts a new vertex of order four
at the centre of a tetrahedron, thereby creating
four new tetrahedra. A $(2,3)$-move takes two tetrahedra
which share a triangle, removes this triangle and replaces
the two tetrahedra by three tetrahedra, whose common axis 
(a link of order three) is dual to the removed triangle.
Both types of moves are shown in Fig.\ \ref{d3move}.
A similar construction in 4d leads to a set of five $(p,q)$-moves.

These finite sets of moves are ergodic, that is,
by repeating local moves, all configurations in the
ensemble can be reached. Unlike in 2d, there are no subsets of moves which
are ergodic in the set of triangulations of fixed volume.
In 3d, all $(p,q)$-moves change the volume. 
If one wants to perform numerical simulations at some fixed $V$,
a modification of the procedure used in the simulations becomes necessary. 

A simple way of forcing the volume to lie close to a given $V$ is
to use a modified action
\beq\label{algorithm}
S_{EH}' = \kappa_d N_d - \kappa_0 N_0 + \epsilon |N_d - V|^{\delta}
\eeq
in the Monte Carlo simulation, typically with 
$\delta =1$ or 2, and a free parameter $\epsilon$.
The Monte Carlo process now takes place in the set of all 
triangulations, with fluctuating 
volume. If we succeed in tuning the cosmological constant to 
criticality,
$\kappa_d \approx \kappa_d^c(\kappa_0)$, the additional term in the
action will force the volume to fluctuate around
$N_d = V$. The amplitude of the fluctuations will depend on the free
parameter $\epsilon$, which should be sufficiently small to permit
big changes in the geometry. We can then simply collect all
configurations with $N_d = V$, and use them in measurements, 
provided they are separated by a number of
steps much bigger than the autocorrelation time.

This method seems straightforward, but it leads to the following
complication. Since our set of moves is ergodic only in the ensemble of 
simplicial manifolds of fixed topology and {\it of arbitrary volume} $N_{d}$,
restricting the possible volume fluctuations {\it could}  make
some parts of the configuration space unreachable.
For manifolds which are not {\it algorithmically recognizable} it can 
be proven \cite{ben-av} (see also \cite{book}) that 
the number of steps needed to connect two arbitrary triangulations of
identical volume $N_d$ cannot be bounded by a computable function $f(N_d)$
(a function which can itself be computed by a finite algorithm).
This situation cannot arise when one has a finite set of moves
which {\it is} ergodic for fixed volume (or even bounded by a 
computable function), since in this case one can just make a list 
of all triangulations, which will have a finite, computable length.
We conclude that 
\begin{itemize}
\item[$\bullet$] {\it if} a manifold is not algorithmically recognizable
there will be ``volume barriers'' in the sense explained above, 
effectively separating parts of the configuration space.
\item[$\bullet$] In a dimension $d$ where algorithmically unrecognizable
manifolds exist, one cannot have a single, finite set of volume-preserving 
moves which at the same time is {\it general}, that is, which works for 
{\it any} manifold of the given dimension, 
and which is ergodic for {\it all} manifolds. 
\end{itemize}
In $d=2$ all manifolds are algorithmically recognizable,
and the flip move (introduced in Section \ref{monte}) is 
a volume-preserving move which is ergodic 
for an arbitrary manifold. It is not known whether there 
exist three-dimensional manifolds which are not algorithmically recognizable.
It has recently been proven that the three-sphere {\it is} algorithmically 
recognizable \cite{thompson}, but it 
is not known whether the same is true for $S^4$. 
For $d>3$ there exist 
algorithmically unrecognizable manifolds, and for dimension $d>4$ 
not even the $d$-sphere is algorithmically recognizable \cite{abb}.

In the 4d numerical simulations we use triangulations
with $S^4$-topology, and the reliability of the 
procedure outlined above could depend on the manifold being
algorithmically recognizable. 
If it is not, one might worry that there 
are configurations, or even large regions 
which are separated from the rest of the 
configuration space by very high ``volume 
barriers''.  
This would imply that in order to reach such a configuration
by a successive application of the elementary moves,
one would be forced to go through very large intermediate volumes.

Attempts have been made to detect such barriers in numerical
simulations on $S^4$, but without success \cite{aj3}. 
Unfortunately, this cannot be taken
as an indication that $S^4$ {\it is} algorithmically recognizable
because even for the five-sphere $S^5$, 
which is known to {\it not} have this property, such barriers
have not been seen \cite{bas}.
There are two ways of interpreting the $S^5$-result. 
Either the Monte Carlo simulation explores
only a fraction of the configuration space (in some sense it does anyway,
since one always deals with a finite sample of ``typical'' configurations),
or the configurations ``beyond the barrier" which are hard to reach 
constitute only a very small fraction
of the configuration space. 
For $S^4$, there need not be any problems
in case it {\it is} algorithmically recognizable. If it is not,  
we hopefully are in a situation where the ``unreachable''
configurations form only a small subset of configurations 
that becomes negligible in the infinite-volume 
limit. 

Just as in 2d, the local moves can be supplemented 
by {\it minbu surgery} moves, using a suitably generalized concept of
(minimal) baby universes. They are again defined as parts of the
triangulation which are connected to the parent universe by 
$(d-1)$-dimensional minimal necks, which topologically are boundaries
of $d$-simplices. 
The baby universes again play an important role, especially in
the large-$\kappa_0$ phase, and using an improved algorithm has 
been essential in
studying the phase structure of the four-dimensional theory. 

The observables described in 2d also exist for gravity 
in higher dimensions.
The minbu distribution can be measured and used to determine
$\gamma_{str}$ (in case the behaviour of $z_V(\kappa_0)$ is power-like),
and the scaling properties of the two-point function $G_V(r)$ 
yield information about the Hausdorff dimension.
Matter degrees of freedom can be included without problems, 
the simplest being spins and massless scalar fields.
The coupling of gauge fields to 4d dynamical triangulations will
be discussed in the next section. -- 
In all these cases we can employ finite-size scaling
and measure the matter correlators.

\subsection{Numerical results in higher dimensions}\label{highdresults}

The algorithm outlined in the previous section was first applied 
to pure 3d gravity \cite{grav3d}. As predicted, both a 
branched-polymer and a crumpled phase were discovered, but the
transition between them turned out to be a very strong first-order
phase transition. This manifests itself in a very large hysteresis
in the order parameter $p(\kappa_0)$, when changing $\kappa_0$ across the
phase transition point. The numerical effort spent on the 3d case was
relatively small. Since the phase transition is first-order, it cannot
give rise to a scaling limit. The phase structure
remained unchanged also when spin fields were added \cite{spin3d}.

Soon afterwards a numerical algorithm for 4d pure gravity was 
constructed \cite{grav4d} and its phase structure 
studied extensively. It looked similar to the one 
in 3d, with two distinct phases. However, this time the phase transition
was much softer, and for a long time was believed to be
continuous. This exciting possibility led to
many numerical explorations of the 
properties of the two phases and the nature of the phase transition.
It was shown that for large $\kappa_0$ one obtains a generic 
branched-polymer phase with $\gamma_{str}=1/2$ and $d_h = 2$. 
Minbu surgery
moves were essential in generating independent configurations in this
phase. The structure of the crumpled phase seemed rather unusual:
the behaviour of $z_V(\kappa_0)$ was no longer dominated by a power
of $V$, and the scaling of $G_V(r)$ suggested that $d_h \to \infty$. 
(This implies that the average linear extension of the system 
is almost independent of $V$.) 
It was then understood that these pathologies are related to the appearance of
exactly two ``singular'' vertices of very large order $\sim V$. 
A finite fraction of the simplices of the triangulations 
shares these two vertices, giving rise to a ``volume condensation''.
The two points are always joined by a link, whose order 
also grows with the volume \cite{singvertices}. 

The appearance of these structures is well 
described both by the mean-field approach mentioned earlier \cite{acm1}, and
by a phenomenological, so-called balls-in-boxes model
\cite{balls}.
This model represents the gravitational system as a collection of $N_0$ boxes, 
where the $i$'th box contains $q_i$ balls, analogous to the order $o(i)$ 
of a vertex $i$ in the simplicial complex.
Note that the vertex orders $o(i)$ on a triangulation satisfy
\beq\label{sumu}
\sum_i o(i) = (d+1) N_d.
\eeq
In the simplified model, the $q_i$ are required to satisfy the same
constraint. 
The local probability of finding $q$ balls in a box is given by
$p(q) \sim q^{-\beta}$, where $\beta$ is a free parameter. 
The model can be solved exactly in the thermodynamic limit, and
exhibits two phases: one in which the balls are distributed randomly in all 
boxes, and a ``crumpled'' phase at large $\beta$ where a finite fraction 
of the balls appears in one box. Note that this model has
no concept of neighbouring boxes, and there are no correlations among
the $q_{i}$ (other than the relation \rf{sumu}).

Although the balls-in-boxes model can hardly claim to reveal much about the
geometric structure of quantum gravity, it nevertheless seems
to capture some important features of
the situation observed in the numerical simulations. 
In particular, one might wonder whether also in the full
simplicial-gravity model there are no correlations between the
vertex orders.
This conjecture can be tested by investigating the corresponding
gravitational two-point correlator for vertices separated by a 
(geodesic) distance $r$. Such a correlator is similar to a matter-matter
correlator with non-vanishing expectation values for the matter fields.
(We will not discuss how to define correctly its connected part.) 
Preliminary
measurements indicate that the correlations do not extend
beyond a few lattice spacings, that is, they are of the order of the cut-off
\cite{AAJgauge} (see also \cite{bas1} for an earlier report).
Similarly short-ranged correlations have also been observed 
in the 2d case \cite{ABJ}. 

\begin{figure}
\centerline{\hbox{\psfig{figure=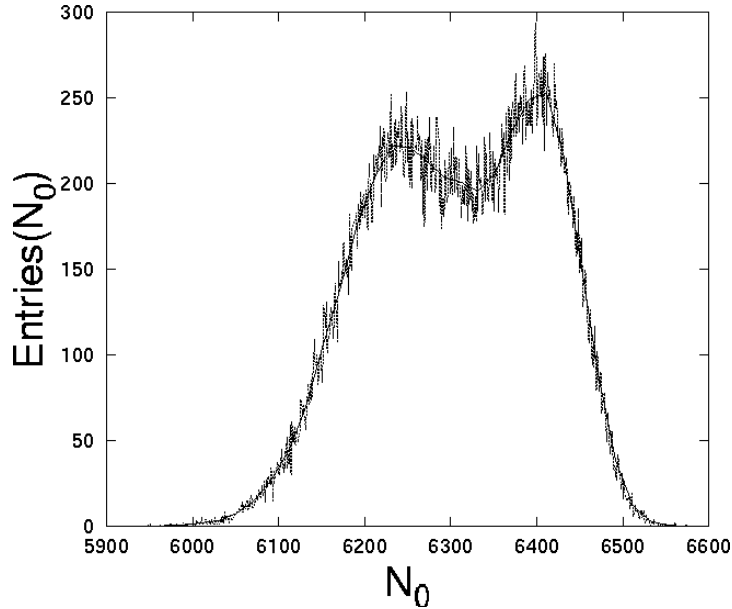,height=8cm,angle=0}}}
\caption[doublep]{The double peak structure at the phase transion 
point as a function of the the number of vertices, $N_0$, for a fixed number
of four-simplices, $N_4\equ 32000$.}
\label{doublep}
\end{figure}

There is by now overwhelming numerical evidence that the
weak- and strong coupling phases are separated by 
a weak first-order transition \cite{andre1}. 
The determination of the order turned out to be difficult, since
no hysteresis was observed for finite systems. 
The best way of identifying the nature of the phase transition is by
measuring the distribution of $N_0$ in a sample of configurations
exactly at the pseudo-critical point. The distribution has
two peaks, whose separation grows with the volume.
This is shown in Fig.\ \ref{doublep}.
The effect becomes only apparent for sufficiently
large systems (of 32000 and 64000 simplices); for smaller
volumes the separation cannot be seen.

From the point of view of quantum gravity, this is of course a
disappointing result, because there is no new continuum physics 
associated with the phase transition.
The weakness of the transition could be an indication that  
a continuous phase transition can be reached by enlarging the
space of coupling constants, that is, by extending the action
$S_{EH}$. Obvious candidates would be terms containing 
higher powers of the curvature, which have previously been used
in quantum Regge calculus \cite{hamber}. 
 
In both 3d and 4d simplicial gravity,
the phases observed on either side of the transition are pathological,
in that they do not seem to correspond to an effective quantum
geometry of a sensible dimensionality.
As already mentioned 
we find either branched polymers (locally one-dimensional objects) or
a condensation of the volume around selected points 
(yielding an object of arbitrarily large dimension). It is 
surprising that the rich geometric structure observed
in two dimensions has all but disappeared. There are potential cures for this,
although there is no guarantee that they will lead to the desired quantum
theory. 

Within the framework of dynamical triangulations, 
there are basically two ways of ``changing the measure", in order
to improve this state of affairs.
One is by using a modified action, but leaving the sum over
geometries untouched. Gravity with a squared scalar-curvature term
was investigated in \cite{charlotte}, 
but the numerical results
indicated that the phase structure remained unchanged. 
Note that also in $d=2$, the addition of such terms does not change 
the universality class of pure gravity, 
as has been shown both numerically (see, for example, \cite{tabor})
and analytically \cite{kazakov}.
The other possibility is to change the sum to a different class of
triangulations. A physically motivated example of this kind 
is the imposition of a causality constraint on Lorentzian geometries.
It {\it does} lead to a change of universality class in $d=2$, as
we have already seen. There are indications that also in $d>2$
a causality condition acts as an effective regulator for the
quantum geometry \cite{underway}. Numerical investigations of Lorentzian
models in higher dimensions are under way.
 
Another logical possibility is that only the matter-coupled
theory possesses a well-defined scaling limit.
Recall that in two dimensions, there is a rather narrow window
of physically interesting Euclidean continuum theories, namely, 
Liouville gravity interacting with conformal
fields of charge $0 \le c \le 1$. 
We simply may have been lucky that pure gravity lies within this range.

Matter interactions could in principle suppress the formation
of baby universes in higher dimensions, 
and the inclusion of gauge (vector) fields 
was suggested as a promising scenario for testing this conjecture  
\cite{ammot}.
The simplest model of this kind is the non-compact 
4d Abelian gauge model formulated in \cite{burdakrz}.
Like on a regular lattice, the gauge
fields $A_{ij}\equiv -A_{ji}$ are associated with oriented 
lattice links $\{i,j\}$.
A discretized version $P$ of the field strength tensor $F_{\m\n}$ 
can be expressed in terms of
``plaquettes" around (oriented) triangles $\{ijk\}$,
\beq\label{plaquette}
P_{\{ ijk\} }= A_{ij}+A_{jk}+A_{ki}.
\eeq  
The action was taken to be
\beq\label{gauge}
S_{gauge}=\sum_{\{ijk\}} o(\{ijk\}) P_{\{ijk\}}^2,
\eeq
with $o(\{ijk\})$ denoting the order of the triangle $\{ijk\}$.
This factor was included
to guarantee that whenever $P^2$ is constant for all triangles,
$S_{gauge}$ is proportional to the volume $N_4$ of the manifold. 
The action has a simple
gauge symmetry and is relatively easy to use in numerical simulations,
since it is Gaussian in the vector fields. 
The number $n_g$ of gauge fields
can be varied, and for $n_g=3$ a new ``crinkled" phase with
$\gamma_{str} < 0$ was observed,
replacing the branched-polymer phase. 
Moreover, the phase transition seemed to be
continuous rather than first-order. 

However, it was soon realized that this new phase can be obtained 
in a model without gauge fields, by adding a geometric term
\beq\label{effective}
S_{\Delta} =  {n_g \over 2} \sum_{\{ijk\}} \log o(\{ijk\})
\eeq
to the pure-gravity action \cite{effect}. Since $S_\Delta$ is an
ultra-local measure term, one would not really expect it to
lead to new {\it continuum} physics.

The origin of this term can be traced to a somewhat unfortunate
choice of gauge action. At first sight, it seems natural to 
associate the gauge fields with the links of the triangulation,
analogous to what is done on regular, flat lattices.
This implies that any charged fields which couple to the 
gauge fields will be located at the vertices. 
However, note that scalar fields coupled to
gravity in all investigations so far have been placed 
at the centres of the four-simplexes, to ensure that the  
number of field degrees of freedom grows proportional to the volume. 
In the case of fluctuating geometry, this need not be the case
if the scalar fields are located at the vertices, since
$N_0$ may grow much slower than $N_4$. 

Thus, if our primary aim is to compare the effects of gauge
fields on geometry with those of the previously studied
scalar fields, the complex-valued charged fields 
should be associated
with the four-simplexes of the triangulation, and the 
gauge fields with the links of the {\it dual} lattice.
The pure-gauge action is then a sum over the {\it dual} plaquettes,
which are the natural duals of the triangles in the original 
triangulation (see \cite{AAJgauge} for details). 

The two models with the gauge potentials $A$ on the links and 
on the dual links are dual to each other. 
The theory defined by \rf{gauge} (on the triangulation)
is equivalent to the gauge theory on the dual lattice,
if we add by hand a term \rf{effective} to the 
natural gauge action on the dual lattice \cite{AAJgauge}. 
Monte Carlo simulations with gauge fields on the dual links
without the term \rf{effective}, 
as advocated above, have not detected any effect of the
gauge fields on the geometry, in line with previous results for 
scalar fields and Ising spins coupled to gravity. 

Putting a more positive spin on this result, the inclusion
of the term \rf{effective} is the only example of a modification 
of the action which has led to a genuine change of the geometry.
Although it has no obvious physical interpretation
and although the crinkled phase is probably not related to interesting 
continuum physics, it does give us some hope of finding a
physically well-motivated term which {\it does} change 
the geometry in an interesting way, allowing us to define 
a continuum theory of gravity.

\section{Outlook}\label{outlook}

In these notes we have described in some detail a non-perturbative
path-integral approach to quantum gravity, with and without matter. 
This ``dynamical
triangulations" method uses an intermediate regularization, in
which the space of all space-time geometries is approximated by
a set of simplicial complexes with certain edge length 
assignments, in the spirit of Regge calculus. For a finite volume
(that is, for a complex with a finite number of simplices) and
fixed space-time topology,
this method yields well-defined, convergent partition functions and
propagators. The aim in this approach is to understand the
critical behaviour of these statistical systems of fluctuating
geometry, and investigate their physical properties in the
continuum limit, if it exists.

We believe that this ansatz has a number of virtues that make
further research worthwhile. Firstly, it works perfectly
in two dimensions, leading to a host of analytical and numerical
results. There, it agrees with other discrete or continuum
formulations, whenever they can be compared, and in some cases
turns out to be even more powerful. This gives us considerable
confidence in the validity of the method. Moreover, the
two-dimensional construction has served as a useful laboratory
for defining diffeomorphism-invariant versions
of two-point functions and other field-theoretic observables.
Dynamical triangulations as a method can be applied in
{\it any} dimension $d$, but obviously will lead to very different
theories, depending on $d$. Only in $d=4$ do we expect to find 
local, interacting quantum-gravitational degrees of freedom. 

Secondly, as has been shown recently, there is also a Lorentzian version 
of the formulation, where additional causality restrictions are
imposed on the space-time geometries, and a well-defined
Wick rotation exists. The absence of a natural notion of
``analytic continuation'' from Euclidean to Lorentzian 
signature has for a 
long time stalled progress in discretized path-integral approaches to 
gravity. Surprisingly, in two space-time dimensions the Euclidean
and Lorentzian continuum quantum gravity theories are very different.
As a consequence of the causality constraint, the ``baby-universe''
structure of the Euclidean theory is absent in the 
Lorentzian sector. This leads to a ``smoother'' quantum geometry, 
and well-behaved matter coupling even beyond the $c=1$ barrier.
Such a behaviour is very desirable in dimension $d>2$, where so far
the dominance of ``extreme'' geometric phases in Euclidean
dynamically triangulated models seems to have prevented the
existence of an interesting continuum limit. 

In higher dimensions, much work remains to be done to establish 
whe\-ther our method leads to a well-defined continuum quantum
theory. In the Euclidean sector, a suitable change of the
discretized measure or the inclusion of different types of matter 
may still lead to long-range correlations. In the Lorentzian
sector, there is a genuine chance that the causality
restrictions will lead to a qualitatively different phase
structure, including a second-order phase transition. The investigation
of the Lorentzian models for $d>2$ has only just begun.
However, we do not have to start from scratch when trying to
solve the theory, but can use the considerable expertise
gathered in the Euclidean simulations. Obviously,
quantum gravity will not be solved by numerical methods alone,
but we have learned  from the example of dynamical triangulations in 
2d that a mathematically exact analysis and numerical
simulations can work hand in hand. 

In addition, one can hope for analytical progress in solving
the discretized models (which in the case of quantum Regge calculus
has always been hampered by the presence of
inequalities on the edge lengths, even in $d=2$). 
In terms of dynamical triangulations, quantum gravity has been
turned into a well-defined combinatorial problem, which has
already been solved exactly in two dimensions. It is also likely
that -- analogous to the Euclidean case -- mean-field methods can
help to elucidate the Lorentzian phase structure in higher dimensions.
This may have interesting ramifications for the general study
of spaces of pseudo-Riemannian geometries.

In summary, the dynamical-triangulations approach 
is a promising tool for both analytical and
numerical investigations of quantum gravity. It allows for a
natural inclusion of matter, and is explicitly 
reparametrization-invariant. The path integral can be defined
for both Euclidean and Lorentz\-ian geometries. In the
Lorentzian case, it involves summing over a class of 
Wick-rotatable space-times with a natural ``time''-slicing,
reminiscent of the geometric structures appearing in 
Hamiltonian quantization schemes. We hope that our
construction of a Lorentzian path integral will serve as a
bridge between the canonical and covariant formulations,
benefitting both and improving our understanding of the nature of
quantum gravity.

\acknowledgements 

J.A.\ acknowledges the support of MaPhySto -- Center for Mathematical Physics 
and Stochastics -- financed by the National Danish Research Foundation. 
J.J.\ was partially supported by the Polish Government KBN grant 
no.\ 2p03B01917.

\end{document}